\documentclass{article}
\usepackage[top=30truemm,bottom=30truemm,left=25truemm,right=25truemm]{geometry}
\usepackage{lscape} 
\usepackage{comment} 

\usepackage[dvipdfmx]{graphicx}

\usepackage{multirow}
\usepackage{graphicx}
\usepackage{bm}
\usepackage{float}
\usepackage{amsmath}
\usepackage{amssymb}
\usepackage{amsthm}

\newtheorem{example}{Example}

\newtheorem{proposition}{Proposition}[section]
\theoremstyle{definition}
\usepackage{amsmath}
\usepackage{amsfonts}
\usepackage[]{graphicx}
\usepackage{booktabs}
\usepackage[ruled,vlined]{algorithm2e}
\chardef\bslash=`\\ 

\begin{document}

\title{Estimation and visualization of \\ treatment effects for multiple outcomes}

\author{Shintaro Yuki\thanks{Graduate School of Culture and Information Science, Doshisha University, Tataramiyakodani 1-3, Kyotanabe City, Kyoto, Japan.}\and
		Kensuke Tanioka\thanks{Department of Life and Medical Sciences, Doshisha University, Tataramiyakodani 1-3, Kyotanabe City, Kyoto, Japan.}  \and
			Hiroshi Yadohisa\thanks{Department of Culture and Information Science, Doshisha University, Tataramiyakodani 1-3, Kyotanabe City, Kyoto, Japan} 
	}

\date{}
\maketitle

%
\begin{abstract}
	We consider a randomized controlled trial between two groups. The objective is to identify a population with characteristics such that the test therapy is more effective than the control therapy. Such a population is called a subgroup. This identification can be made by estimating the treatment effect and identifying interactions between treatments and covariates. To date, many methods have been proposed to identify subgroups for a single outcome. There are also multiple outcomes, but they are difficult to interpret and cannot be applied to outcomes other than continuous values. In this paper, we propose a multivariate regression method that introduces latent variables to estimate the treatment effect on multiple outcomes simultaneously. The proposed method introduces latent variables and adds Lasso sparsity constraints to the estimated loadings to facilitate the interpretation of the relationship between outcomes and covariates.  The framework of the generalized linear model makes it applicable to various types of outcomes. Interpretation of subgroups is made by visualizing treatment effects and latent variables. 
This allows us to identify subgroups with characteristics that make the test therapy more effective for each multiple outcomes. Simulation and real data examples demonstrate the effectiveness of the proposed method.
\end{abstract}


\section{Introduction}
\label{s:intro}
Clinical trials are conducted to compare the efficacy of a test therapy with that of a control therapy, and this paper, deals specifically with two-arm randomized controlled trials.
In a two-arm randomized controlled trial, subjects are randomly assigned to the test or control therapy to compare the effectiveness of the test treatment.
However, the results of clinical trials may not demonstrate the efficacy of the test therapy in people that meet the eligibility criteria.
In such cases, it is desirable to efficiently detect populations that have characteristics that would make the test therapy more effective than the control therapy, see Lipkovich, Dmitrienko, and B.D'Agostino Sr (2017), Pocock et al. (2002), Svetkey (1999), Yusuf (1991) and, Bonetti and Gelber (2004).
They can be detected by estimating the treatment effect (Rubin, 1974) and identifying the interaction between the treatment and the covariates.
If we try to calculate the treatment effect directly, we need both the subject's results from the control therapy and the test therapy. However, since subjects only receive one therapy, either the test or the control therapy, one of the results is missing.
Therefore, it is necessary to estimate the treatment effect from the data of one treatment for each subject.
\par
Today, there are many models for estimating treatment effects, varying from randomized controlled trials (e.g., Twisk, 2018; Imbens and Rubin, 2015) to observational studies (e.g., Lunceford and Davidian, 2004; Athey and Imbens, 2017).
We now present specific examples of ways to estimate treatment effects and identify interactions between treatments and covariates when assuming a randomized controlled trial.
One of them is the linear regression model, in which the product of a binary treatment index and a baseline covariate is included in the regression model, to examine the interaction between the treatment and the covariate.
Gustafson (2000) proposed a spline-based method for flexible Bayesian regression that models smooth bivariate interactions.
Elsewhere, Bonetti and Gelber (2004) proposed a subpopulation treatment effect pattern plots: STEPP, to examine the relationship between treatment effects and these corresponding covariates.
In addition to that, Bonetti et al. (2009) and Sauerbrei, Royston, and Zapien (2007) developed and modified STEPP. 
Most of these existing models are not designed for the analysis of high-dimensional covariates.
Therefore, Tian et al. (2014) proposed a modified outcome method: MOM to estimate the treatment effect using Lasso (Tibshirani, 1996) for high-dimensional covariates.
Although MOM is a linear regression model for estimating the treatment effect on a single outcome, Guo et al. (2021) proposed the multiple outcomes treatment effect forests: MOTEF, an extension of MOM that can be applied to multiple outcomes.
However, MOTEF has been proposed only for continuous outcomes.
The outcomes may be binary when estimating the treatment effect (e.g., Rosenbaum and Rubin, 1983; Hu et al., 1998; Horowits and Manski, 2000).
Moreover, there are many opportunities to use multi-level data in clinical trials (e.g., Griswold, Localio, and Mulrow, 2010).
In addition, since MOTEF is based on the random forests, it has difficulty in interpretation of subgroup.
To overcome these problems of MOTEF, we propose the new approach for various types of multiple outcomes with latent variables.
\par
As one of the regression methods with latent variables,
sparse principal component regression: SPCR
has been proposed by Kawano et al. (2015, 2018) although this method 
is for a single outcome, not for multiple outcomes. In this method, the principal components that contribute to the regression analysis can be estimated by simultaneously performing sparse principal component analysis and a generalized linear model based on Lasso.
A method is also proposed for multi-block data with a similar objective function (Gvaladze et al., 2021; Van Deun, Crompvoets, and Ceulemans, 2018; Park, Ceulemans, and Van Deun, 2021).
However, these methods are for single outcome and have not been used in a framework for estimating treatment effects.
\par
In this paper, we propose a method that is easy to discover and interpret subgroups and can be applied to various types of outcomes.
Specifically, to discover subgroups, components that have a common and significant impact on treatment effects, are found through exploration.
In addition to that, sparse constraints by Lasso are added to the estimated coefficient matrix and loading matrix.
By visualizing these relationships in a path diagram, treatment effects can be easily interpreted, and subgroups can be made based on them.
The method is different from the two-stage procedure, in which common components are found by principal component analysis: PCA (Jolliffe, 1986), and regression analysis is performed on the outcome using these components, because it enables us to adaptively provide sparse principal component loadings that are associated with the outcomes based on SPCR.
\par
This paper demonstrates the usefulness of the proposed method, and the validity of the discovered subgroups through simulations and real data analysis.
This paper is organized as follows. 
In Section 2, we describe the model of the outcome and the introduction of latent variables, and show the proposed method.
In Section 3, we describe the objective function of the proposed method for multiple continuous outcomes, and the algorithm for estimating the parameters.
In Section 4, we describe the objective function of the proposed method for multiple binary outcomes, and the algorithm for estimating the parameters.
In Section 5, we generalize the proposed method for various types of outcomes, reduction methods and penalty functions.
In Section 6, we perform numerical simulations of the proposed method.
In Section 7, we apply the proposed method to the real data.
We conclude in Section 8.
\section{Model of outcome}
In this section, we show the notations to describe the proposed model at first. Subsequently, we propose the extension of the MOM to the framework of multiple outcomes. The proposed model can estimate the treatment effects for the multiple outcomes of multiple subjects. MOM is an existing model for estimating the treatment effect. However, this can only be applied to the univariate outcomes, and no model for estimating treatment effects for the multiple outcomes of multiple subjects has been proposed. Therefore, we describe the proposed model in two stages. The first step is extending modified outcome method to the case of the multiple outcomes of a single subject.
The second step is extending modified outcome method to the case of the multiple outcomes of multiple subjects.
Next, we introduce basic notations. Let $T_i$ be a random variable as follows, 
\begin{align}\nonumber
    T_i= \left\{
\begin{array}{ll}
1 & (\mathrm{If\ subject\ \textit{i}\ is\ allocated\ test\ therapy}) \\
-1 & (\mathrm{If\ subject\ \textit{i}\ is\ allocated\ control\ therapy})
\end{array}
\right. 
\end{align}
and $t_{i}$ be an observed value of $T_i$.
We consider a randomized clinical trial in this case, i.e., $P(T_i=1)=P(T_i=-1)=0.5$. 
Let $\boldsymbol{T}=\mathrm{diag}(T_{1},T_{2},\cdots,T_{n}),\boldsymbol{T}^{(\mathrm{obs})}=\mathrm{diag}(t_{1},t_{2},\cdots,t_{n})$ be a binary treatment indicator matrix.
In extending MOM, 
$Y_{i}=(Y_{i1},Y_{i2},\cdots,Y_{ip})^{\prime}\in\mathbb{R}^{p}$ is the multiple vector of random variable corresponding to the outcome of subject $i$. 
Here, treatment effects are defined as the expected value of the difference in potential outcomes (Rosenbaum and Rubin, 1983) when $Y_{i}$ is continuous, 
and are defined as the log odds ratio when $Y_{i}$ is binary. 
$\boldsymbol{X}=(\boldsymbol{x}_{1},\boldsymbol{x}_{2},\cdots,\boldsymbol{x}_{n})^{\prime}=(\boldsymbol{1}_n,\boldsymbol{x}_{(1)},\boldsymbol{x}_{(2)},\cdots,\boldsymbol{x}_{(m)})\in\mathbb{R}^{n\times (m+1)}$ is an explanatory variable matrix with $n$ subjects and $m$ variables, $\boldsymbol{x}_{i}=(1,x_{i1},x_{i2},\cdots,x_{im})^{\prime}$ for the subject $i$ and $\boldsymbol{x}_{(j)}=(x_{1j},x_{2j},\cdots,x_{nj})^{\prime}$ for the variable $j$, and $\boldsymbol{1}_{n}$ is the $n$-dimensional vector whose elements are all one.
We assume that treatment effects can be expressed as a linear function of explanatory variables.\par
First, we introduce the simple regression model for multiple continuous outcomes. We consider the multiple linear regression model to estimate treatment effects of subject $i$ as follows,
\begin{align}\label{full1}
    Y_{i}=f(\boldsymbol{x}_{i})+\frac{T_{i}}{2}\boldsymbol{\varGamma }^{\prime}\boldsymbol{x}_{i}+\boldsymbol{\varepsilon}_{i}
\end{align}
where $f:\mathbb{R}^m\mapsto \mathbb{R}^{p}$ is the arbitrary function for the main effect term, $\boldsymbol{\varGamma}=(\boldsymbol{\gamma}_{1},\boldsymbol{\gamma}_{2},\cdots,\boldsymbol{\gamma}_{p})\in\mathbb{R}^{(m+1)\times p}$ is the regression coefficient matrix for the interaction term and $\boldsymbol{\varepsilon}_{i}$ is a vector of random errors for subject $i$ and
it is assumed to be independent of $T_i$, i.e., $\boldsymbol{\varepsilon}_i\perp T_{i}$, and $\boldsymbol{\varepsilon}_i\sim P$ with $\mathbb{E}[\boldsymbol{\varepsilon}_{i}]=\boldsymbol{0}_{p}$, where $\boldsymbol{0}_{p}$ is the $p$-dimensional vector whose elements are all zeros. 
\par
Let $Y_i^{(1)}$ be $Y_i$ if subject $i$ is allocated experimental therapy, and $Y_i^{(-1)}$ be $Y_i$ if subject $i$ is allocated control therapy.
These are called potential outcomes and formulated as follows,
\begin{align}
    \nonumber
    Y_{i}^{(1)}&=f(\boldsymbol{x}_{i})+\frac{1}{2}\boldsymbol{\varGamma }^{\prime}\boldsymbol{x}_{i}+\boldsymbol{\varepsilon}_{i}^{(1)},\quad 
    Y_{i}^{(-1)}=f(\boldsymbol{x}_{i})-\frac{1}{2}\boldsymbol{\varGamma }^{\prime}\boldsymbol{x}_{i}+\boldsymbol{\varepsilon}_{i}^{(-1)}
\end{align}
where $\boldsymbol{\varepsilon}_{i}^{(1)},\boldsymbol{\varepsilon}_{i}^{(-1)}\sim P$ with $\mathbb{E}[\boldsymbol{\varepsilon}_{i}^{(1)}]=\mathbb{E}[\boldsymbol{\varepsilon}_{i}^{(-1)}]=\boldsymbol{0}_p$.
Under this assumption, treatment effects can be described as follows,
\begin{align*}
    \mathbb{E}[Y_{i}^{(1)}-Y_{i}^{(-1)}]=\boldsymbol{\varGamma }^{\prime}\boldsymbol{x}_{i}.
\end{align*}\par
Thus, the treatment effects satisfy the assumption that they can be expressed as a linear function of explanatory variables. In this model, the key is that $T_{i}$ takes a binary value of either $-1$ or $1$. This allows us to estimate the treatment effects without considering the main effect term.
\par
On the other hand, the expected value of $2T_{i}Y_{i}$ in randomized clinical trials can be described as follows,
\begin{align*}
    \mathbb{E}[2T_{i}Y_{i}]
    &=\boldsymbol{\varGamma}^{\prime}\boldsymbol{x}_{i}.
\end{align*}
where $\mathbb{E}[T_{i}]=0$ and $\mathbb{E}[T_{i}^{2}]=1$.
\par
That is to say, the expected value of $2T_{i}Y_{i}$ in randomized clinical trials has the same value of treatment effects.
Let $\boldsymbol{\varepsilon}^{*}_{i}$ be a vector of random error of subject $i$. 
We assume $\boldsymbol{\varepsilon}^{*}_{i}\perp T_{i}$ and $\boldsymbol{\varepsilon}^{*}_{i}\sim P^{*}$ with $\mathbb{E}[\boldsymbol{\varepsilon}^{*}_{i}]=\boldsymbol{0}_{p}$. 
Then, instead of $Y_i^{(1)}-Y_i^{(-1)}$, we model $2T_iY_i$ as follows,
\begin{align*}
    2T_{i}Y_{i}=\boldsymbol{\varGamma}^{\prime}\boldsymbol{x}_{i}+\boldsymbol{\varepsilon}^{*}_{i}.
    \end{align*}
    When $T_i$ is observed as $t_i$, $Y_i$ can be modeled as follows,
    \begin{align}
       \label{conti_ou} Y_{i}=\frac{t_i}{2}\boldsymbol{\varGamma}^{\prime}\boldsymbol{x}_{i}+\boldsymbol{\varepsilon}^{\dagger}_{i},
    \end{align}
    where $\boldsymbol{\varepsilon}^{\dagger}_{i}\sim P^{\dagger}$ with $\mathbb{E}[\boldsymbol{\varepsilon}^{\dagger}_{i}]=\boldsymbol{0}_p$.
    \subsection{Multiple continuous outcomes of single subject}
    In this subsection, when $Y_i$ is continuous, we consider a multivariate normal distribution.
    With assuming $\boldsymbol{\varepsilon}_i^{\dagger}\sim N(\boldsymbol{0}_p,\boldsymbol{\Sigma})$ in Eq. $(\ref{conti_ou})$, the distribution of $Y_i$ is represented as follows
    \begin{align*}
        P(Y_{i}=\boldsymbol{y}_{i}|\boldsymbol{x}_i,T_i=t_i)=\frac{1}{\sqrt{(2\pi)^{p}|\boldsymbol{\Sigma}|}}\exp\left\{-\frac{1}{2}\left(\boldsymbol{y}_i-\frac{t_i}{2}\boldsymbol{\varGamma}^{\prime}\boldsymbol{x}_{i}\right)^\prime\boldsymbol{\Sigma}^{-1}\left(\boldsymbol{y}_i-\frac{t_i}{2}\boldsymbol{\varGamma}^{\prime}\boldsymbol{x}_{i}\right)\right\}.
    \end{align*}
    \par
Since Eq. ($\mathrm{\ref{full1}}$) is a linear model, the prediction accuracy of treatment effects worsens when the true main effect term is nonlinear, e.g., $Y_{i}=f(\boldsymbol{x}_{i})+T_{i}/2\cdot \boldsymbol{\varGamma}^{\prime}\boldsymbol{x}_{i}+\boldsymbol{\varepsilon}_{i}$ where $f:\mathbb{R}^{m}\mapsto\mathbb{R}^{p}$ be nonlinear function. Moreover, as the number of parameters to be estimated increases, the prediction accuracy gets worse. However, modified outcome method does not need to estimate the main effect term. This is the advantage of this model.
The parameter $\boldsymbol{\varGamma}$ can be estimated by the least squares method using Eq. $(\ref{conti_ou})$.
\subsection{Multiple binary outcomes of single subject}
In this subsection, we propose MOM that considers the multiple outcomes when $Y_i$ is binary.

\par
Here, we consider logistic regression model, then $Y_{ij}$, i.e., $j$-th element of $Y_{i}$ in Eq. $(\ref{conti_ou})$, can be modeled below as binary random variable, using the fact the $t_i=1/t_i$ since $t_i$ takes a value of either $-1$ or $1$.
\begin{align*}
    Y_{ij}=\frac{t_{i}}{2}\boldsymbol{\gamma}_{j}^{\prime}\boldsymbol{x}_{i}+\varepsilon^\dagger_{ij}.
\end{align*}
where $\varepsilon_{ij}^\dagger \sim P^\dagger$ with $\mathbb{E}[\varepsilon_{ij}^\dagger]=0$, $P^\dagger$ denotes the standard logistic distribution.
\par
Let $P(Y^{(1)}_{ij}=1|\boldsymbol{x}_i)$ be the probability of $Y_{ij}^{(1)}=1$ conditional on $\boldsymbol{x}_i,T_i=1$, and $P(Y^{(-1)}_{ij}=1|\boldsymbol{x}_i)$ be the probability of $Y_{ij}^{(-1)}=1$ conditional on $\boldsymbol{x}_i,T_i=-1$.
In the framework of logistic regression, the link function is the log odds ratio, which models the expected value of $Y_{ij}$.
Thus
\begin{align*}
    &\mathbb{E}[Y^{(1)}_{ij}]=\log \frac{P(Y_{ij}^{(1)}=1|\boldsymbol{x}_i)}{1-P(Y_{ij}^{(1)}=1|\boldsymbol{x}_i)}\;\Longleftrightarrow\; P(Y_{ij}^{(1)}=1|\boldsymbol{x}_i)=
    \frac{\mathrm{exp}( \boldsymbol{\gamma}_{j}^{\prime}\boldsymbol{x}_{i}/2)}{1+\mathrm{exp}( \boldsymbol{\gamma}_{j}^{\prime}\boldsymbol{x}_{i}/2)}.
\end{align*}
In the same way, $P(Y_{ij}^{(-1)}=1|\boldsymbol{x}_i)$ can be expressed as follows,
\begin{align*}
    P(Y_{ij}^{(-1)}=1|\boldsymbol{x}_i)=
    \frac{\mathrm{exp}(-\boldsymbol{\gamma}_{j}^{\prime}\boldsymbol{x}_{i}/2)}{1+\mathrm{exp}(-\boldsymbol{\gamma}_{j}^{\prime}\boldsymbol{x}_{i}/2)}.
\end{align*}
Based on the above, the distribution of $Y_{ij}$ is represented as follows,
\begin{align}
    \label{mc}
 P(Y_{ij}=1|\boldsymbol{x}_{i},T_i=t_i)=\frac{\mathrm{exp}(t_{i}/2\cdot \boldsymbol{\gamma}_{j}^{\prime}\boldsymbol{x}_{i})}{1+\mathrm{exp}(t_{i}/2\cdot \boldsymbol{\gamma}_{j}^{\prime}\boldsymbol{x}_{i})}.
    \end{align}
The parameter $\boldsymbol{\gamma}_{j}$ can be estimated by the maximum likelihood method using Eq. $(\ref{mc})$.
\subsection{Multiple outcomes of single subject with latent variables}
In this subsection, we propose models of multiple outcomes with latent variables.
Here, we introduce a latent variable for the treatment effects.
This allows us to find components that have a common influence on treatment effects from covariates.
\par
From the previous description, we know that the treatment effect of $Y_i$ is described as $\boldsymbol{\varGamma}^\prime\boldsymbol{x}_i$.
We propose a new model with latent variables as follows,
\begin{align}\label{late1}
    &Y_i=\frac{t_i}{2}\boldsymbol{\varGamma}^{*^\prime}\boldsymbol{A}^\prime\boldsymbol{x}_i+\Tilde{\boldsymbol{\varepsilon}}_i
    \quad\mathrm{subject\ to}\  \boldsymbol{A}^\prime\boldsymbol{A}=\boldsymbol{I}_d
\end{align}
where $\boldsymbol{I}_d\in\mathbb{R}^{d\times d}$ is identity matrix, $\boldsymbol{A}=(\boldsymbol{a}_{1},\boldsymbol{a}_{2},\cdots,\boldsymbol{a}_{d})\in\mathbb{R}^{(m+1)\times d}$ is loading matrix with column orthogonal constraint, and $d\leq m$. 
$\boldsymbol{\varGamma}^{*}=(\boldsymbol{\gamma}^{*}_{1},\boldsymbol{\gamma}^{*}_{2},\cdots,\boldsymbol{\gamma}^{*}_{p})\in\mathbb{R}^{d\times p}$ is the coefficient matrix for latent variables $\boldsymbol{A}^\prime \boldsymbol{x}_i$ and $\Tilde{\boldsymbol{\varepsilon}}_i\sim \Tilde{P}$ with $\mathbb{E}[\Tilde{\boldsymbol{\varepsilon}}_i]=\boldsymbol{0}_p$.
\par
When $Y_i$ is continuous, we introduce latent variables into Eq. $(\mathrm{\ref{conti_ou}})$, and the result is as follows, 
\begin{align}\label{smr-1}
        P(Y_{i}=\boldsymbol{y}_{i}|\boldsymbol{x}_i,T_i=t_i)=\frac{1}{\sqrt{(2\pi)^{p}|\boldsymbol{\Sigma}|}}\exp\left\{-\frac{1}{2}\left(\boldsymbol{y}_i-\frac{t_i}{2}\boldsymbol{\varGamma}^{*^\prime}\boldsymbol{A}^\prime\boldsymbol{x}_i\right)^\prime\boldsymbol{\Sigma}^{-1}\left(\boldsymbol{y}_i-\frac{t_i}{2}\boldsymbol{\varGamma}^{*^\prime}\boldsymbol{A}^\prime\boldsymbol{x}_i\right)\right\}.
    \end{align}
\par
\par
When $Y_i$ is binary, in the similar way of continuous case, we introduce latent variables into Eq. $(\ref{mc})$, and the result is as follows,
\begin{align}
    \label{mc2nn}
 P(Y_{ij}=1|\boldsymbol{x}_{i},T_i=t_i)=\frac{\mathrm{exp}(t_{i}\,\boldsymbol{\gamma}_{j}^{*^\prime}\boldsymbol{A}^\prime\boldsymbol{x}_{i}/2)}{1+\mathrm{exp}(t_{i}\,\boldsymbol{\gamma}_{j}^{*^\prime}\boldsymbol{A}^\prime\boldsymbol{x}_{i}/2)}.
    \end{align}
    This means that we reduce the dimension of covariates by using $\boldsymbol{A}$.
The advantage of introducing latent variables is that it simplifies the process of interpretation.    This is because, we reduce the rank of the covariates and assume a latent low-dimensional structure, and the estimated loading matrix is column-orthogonal, which means that each pair of principal components is uncorrelated.
\subsection{Multiple outcomes of multiple subjects}\label{subs}
In this subsection, we show the proposed models for continuous and binary outcomes of multiple subjects as Eq. $(\ref{late1})$ and $(\ref{mc2nn})$, respectively.
Let $\boldsymbol{Y}=(Y_1,Y_2,\cdots,Y_n)^\prime=(Y_{(1)},Y_{(2)},\cdots,Y_{(p)}) \in\mathbb{R}^{n\times p}$ be the matrix of random variable corresponding to the outcome, $Y_{(j)}=(Y_{1j},Y_{2j},\cdots,Y_{nj})^\prime$ for the variable $j$.
\par
The model of multiple outcomes of multiple subjects is described as follows, 
\begin{align}\label{smr}
    \boldsymbol{Y}
    =\frac{1}{2}\boldsymbol{T}^{(\mathrm{obs})}\boldsymbol{XA\varGamma}^{*}+\Tilde{\boldsymbol{E}}
\end{align}
where $\Tilde{\boldsymbol{E}}=(\Tilde{\boldsymbol{\varepsilon}}_1,\Tilde{\boldsymbol{\varepsilon}}_2,\cdots,\Tilde{\boldsymbol{\varepsilon}}_n)^\prime$ is the matrix of random errors. 
Since $\mathbb{E}[\Tilde{\boldsymbol{\varepsilon}}_i]=\boldsymbol{0}_p$, then $\mathbb{E}[\Tilde{\boldsymbol{E}}]=\boldsymbol{O}_{n\times p}$, where $\boldsymbol{O}_{n\times p}$ is the matrix whose elements are all zeros.
We have named this model as the Structured Multiple Regression with Modified Outcome Method: SMR-MOM.
\par
When there are multiple cases, the conditions of $\boldsymbol{x}_i$ and $T_i=t_i$ in Eq. $(\ref{smr-1})$ change to that of $\boldsymbol{X}$ and $\boldsymbol{T}=\boldsymbol{T}^{(\mathrm{obs})}$, resulting in the following distribution,
\begin{align*}
       P(Y_{i}=\boldsymbol{y}_{i}|\boldsymbol{X},\boldsymbol{T}=\boldsymbol{T}^{(\mathrm{obs})})=\frac{1}{\sqrt{(2\pi)^{p}|\boldsymbol{\Sigma}|}}\exp\left\{-\frac{1}{2}\left(\boldsymbol{y}_i-\frac{t_i}{2}\boldsymbol{\varGamma}^{*^\prime}\boldsymbol{A}^\prime\boldsymbol{x}_i\right)^\prime\boldsymbol{\Sigma}^{-1}\left(\boldsymbol{y}_i-\frac{t_i}{2}\boldsymbol{\varGamma}^{*^\prime}\boldsymbol{A}^\prime\boldsymbol{x}_i\right)\right\}.
    \end{align*}
\par
\begin{figure}
[h]
\begin{center}
\includegraphics[width=10cm]{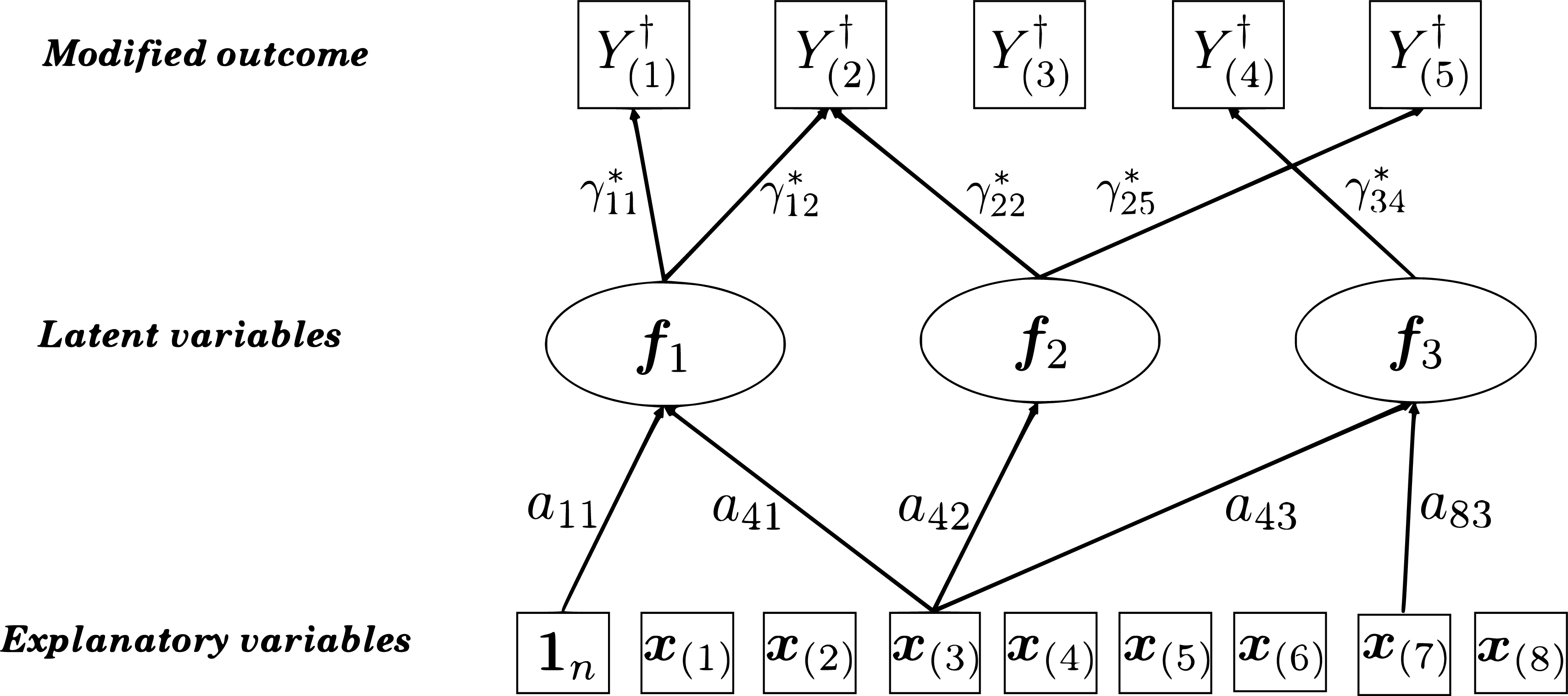}
\caption{Path diagram how the treatment effect is estimated for multiple continuous outcomes, where $Y^\dagger_{(l)}=2\boldsymbol{T}^{(\mathrm{obs})}Y_{(l)}\ (l=1,2,\cdots,p)$}
\label{fig2}
\end{center}
\end{figure}

In the same way, Eq. $(\ref{mc2nn})$ is rewritten as the following distribution,
\begin{align}\label{mc2n}
     P(Y_{ij}=1|\boldsymbol{X},\boldsymbol{T}=\boldsymbol{T}^{(\mathrm{obs})})=\frac{\mathrm{exp}(t_{i}\boldsymbol{\gamma}_{j}^{*^\prime}\boldsymbol{A}^\prime\boldsymbol{x}_{i}/2)}{1+\mathrm{exp}(t_{i}\boldsymbol{\gamma}_{j}^{*^\prime}\boldsymbol{A}^\prime\boldsymbol{x}_{i}/2)}.
\end{align}
For the parameter estimation of these two proposed models, we impose sparsity constraints on both loading of covariates $\boldsymbol{A}$, and the coefficient matrix for latent variables $\boldsymbol{\varGamma}^*$.
The following is an example of the estimation of treatment effects using SMR-MOM for multiple continuous outcomes of multiple subjects and its visualization.
\begin{example}
Here, we show in the path diagram how the treatment effect is estimated by SMR-MOM when the outcome is continuous.
Let latent variables be $\boldsymbol{F}\in\mathbb{R}^{(m+1)\times d}$, i.e.,  $\boldsymbol{F}=(\boldsymbol{f}_{1},\boldsymbol{f}_{2},\cdots,\boldsymbol{f}_{d})=(\boldsymbol{Xa}_{1},\boldsymbol{Xa}_{2},\cdots,\boldsymbol{Xa}_{d})$.
Let $m=8,p=5$ and $d=3$. And suppose that $\boldsymbol{A}$ and $\boldsymbol{\varGamma}^{*}$ are estimated as follows,
\begin{align}\nonumber
\underset{9\times 3}{\boldsymbol{A}}=
    \begin{pmatrix}
a_{11}&0&0&a_{41}&0&0&0&0&0\\
0&0&0&a_{42}&0&0&0&0&0\\
0&0&0&a_{43}&0&0&0&a_{83}&0
    \end{pmatrix}^\prime,\quad
    \underset{3\times 5}{\boldsymbol{\varGamma}^{*}}=
    \begin{pmatrix}
    \gamma^{*}_{11}&\gamma^{*}_{12}&0&0&0\\
    0&\gamma^{*}_{22}&0&0&\gamma^{*}_{25}\\
    0&0&0&\gamma^{*}_{34}&0
    \end{pmatrix}.
\end{align}
In this case, the treatment effect can be described on the path diagram shown in Fig. $\ref{fig2}$.
\end{example}
\section{Objective function and algorithm of SMR-MOM for continuous case}
In this section, we present the objective function of SMR-MOM for multiple continuous outcomes of multiple subjects, and its algorithm for solving optimization problems. 
In addition, we use Lasso estimate for the loading matrix and the coefficient matrix, to ease the interpretation of latent variables and their influence on treatment effects.
In the objective function of SMR-MOM, the framework of sparse principal component regression: SPCR (Kawano et al., 2015) will be used.
\subsection{Objective function of SMR-MOM for continuous case}\label{aaa}
When the multiple outcomes are continuous, the objective function of SMR-MOM is formulated as the following least squares problem:
\begin{align}\label{obj}
    &\underset{\boldsymbol{A},\boldsymbol{B},\boldsymbol{\varGamma}^{*}}{\mathrm{min}}
    \left\{\|\boldsymbol{Y}-\frac{1}{2}\boldsymbol{T}^{(\mathrm{obs})}\boldsymbol{XA\varGamma}^{*}\|^{2}_{F}+\omega\|\boldsymbol{X}-\boldsymbol{XAB}^{\prime}\|^{2}_{F}+\lambda_{a}\sum_{k=1}^{d}\|\boldsymbol{a}_{k}\|_{1}+\lambda_{\gamma^*}\sum_{l=1}^{p}\|\boldsymbol{\gamma}^{*}_{l}\|_{1}\right\},
    \\ \nonumber
    &\qquad\qquad\mathrm{subject\ to\ }\boldsymbol{B}^{\prime}\boldsymbol{B}=\boldsymbol{I},
\end{align}
where $\omega$ is a positive tuning parameter for dimensional reduction, and $\boldsymbol{B}=(\boldsymbol{b}_{1},\boldsymbol{b}_{2},\cdots,\boldsymbol{b}_{d})\in\mathbb{R}^{(m+1)\times d}$ is loading matrix for the latent variables with column orthogonal constraint, and $\lambda_{a},\lambda_{\gamma^*}(\geq0)$ are regularization parameters. 
$\|\cdot\|_F$ is the Frobenius norm, and $\|\cdot\|_1$ is the $L_1$ norm.
The first term of this objective function is squared loss of a linear regression model, and the second term is the loss function of PCA, and the others are the $L_{1}$-regularized terms.
This is the extension of SPCR to the multiple outcomes.
\par
Now, by making the appropriate assumption on $\boldsymbol{Y}$, the following holds.
\begin{align*}
    \underset{\boldsymbol{A},\boldsymbol{\varGamma}^*}{\mathrm{argmin}}\|\boldsymbol{Y}-\frac{1}{2}\boldsymbol{T}^{(\mathrm{obs})}\boldsymbol{XA\varGamma}^{*}\|^{2}_{F}=
    \underset{\boldsymbol{A},\boldsymbol{\varGamma}^*}{\mathrm{argmax}}\;\ell_c(\boldsymbol{A},\boldsymbol{\varGamma}^*),
\end{align*}
where $\ell_c (\boldsymbol{A},\boldsymbol{\varGamma}^*)$ is log-likelihood function of $\boldsymbol{Y}$, and it follows that 
\begin{align*}
    \ell_c (\boldsymbol{A},\boldsymbol{\varGamma}^*)
    =\sum_{i=1}^n \log P(Y_i=\boldsymbol{y}_i|\boldsymbol{X},\boldsymbol{T}=\boldsymbol{T}^{(\mathrm{obs})}).
\end{align*}
Therefore, the objective function of SMR-MOM is newly described as follows:
\begin{align*}
    &\underset{\boldsymbol{A},\boldsymbol{B},\boldsymbol{\varGamma}^{*}}{\mathrm{min}}
    \left\{-\ell_c(\boldsymbol{A},\boldsymbol{\varGamma}^*)+\omega\|\boldsymbol{X}-\boldsymbol{XAB}^{\prime}\|^{2}_{F}+\lambda_{a}\sum_{k=1}^{d}\|\boldsymbol{a}_{k}\|_{1}+\lambda_{\gamma^*}\sum_{l=1}^{p}\|\boldsymbol{\gamma}^{*}_{l}\|_{1}\right\},
    \\ \nonumber
    &\qquad\qquad\mathrm{subject\ to\ }\boldsymbol{B}^{\prime}\boldsymbol{B}=\boldsymbol{I}.
\end{align*}
\subsection{Algorithm of SMR-MOM for continuous case}
In this subsection, we present our algorithm for solving the optimization problems and demonstrate how the parameters are updated. Each parameter is updated by the alternating least squares method: ALS (Young et al., 1980). Since penalty terms exist,
the parameters $\boldsymbol{A}$ and $\boldsymbol{\varGamma}^{*}$ are updated by the proximal gradient method (Rockafellar, 2015). The parameter $\boldsymbol{B}$, which exists only in the third term of Eq. $(\ref{obj})$ and has orthogonal constraints, is updated by singular value decomposition.
\par
\begin{algorithm}[H]\label{al1}
\SetAlgoLined
\KwResult{$\boldsymbol{A},\boldsymbol{B},\boldsymbol{\varGamma}^{*}$}
 \textit{initialize} $\boldsymbol{A},\boldsymbol{B},\boldsymbol{\varGamma}^{*}$\\
 \While{the objective function converges}{
  \textit{Update $\boldsymbol{A}$ using proximal gradient method (by Eq. (\ref{soft}))}\\
  \textit{Update $\boldsymbol{B}$ using singular value decomposition (by Eq. (\ref{BBB}))}\\
  \textit{Update $\boldsymbol{\varGamma}^{*}$ using proximal gradient method (by Eq. (\ref{soft2}))}\\
 }
 \caption{SMR-MOM for continuous case}
\end{algorithm}
In order to estimate these parameters, we use the following propositions.
\begin{proposition}
Given $\boldsymbol{B}$ and $\boldsymbol{\varGamma}^{*}$, $\boldsymbol{A}$ is updated to $\widehat{\boldsymbol{A}}$ in order to minimize the objective function by the proximal gradient method;
    \begin{align}\label{soft}
    \widehat{a}_{ij}&\leftarrow \left\{
\begin{array}{ll}
a_{ij}^{\dagger}-\lambda_{a} & (a_{ij}^{\dagger}>0\ \textit{and}\ \lambda_{a}<|a_{ij}^{\dagger}|) \\
a_{ij}^{\dagger}+\lambda_{a}& (a_{ij}^{\dagger}<0\ \textit{and}\ \lambda_{a}<|a_{ij}^{\dagger}|)\\
0&(\lambda_{a}\geq |a_{ij}^{\dagger}|)\quad\quad\quad\quad
\end{array}
\right.,
\end{align}
where $a_{ij}^{\dagger}$ denotes the $(i,j)$ elements of
\begin{align}
\nonumber
    \boldsymbol{A}^{\dagger}&=\boldsymbol{A}-\alpha_{a}\left\{\left(-\boldsymbol{X}^{\prime}\boldsymbol{T}^{(\mathrm{obs})}\bm{Y\varGamma}^{*^\prime}+\frac{1}{2}\boldsymbol{X}^{\prime}\boldsymbol{XA\varGamma}^{*}\boldsymbol{\varGamma}^{*^\prime}\right)+\omega\left(-2\boldsymbol{X}^{\prime}\boldsymbol{XB}+2\boldsymbol{X}^{\prime}\boldsymbol{XA}\right)\right\},
\end{align}
and $\alpha_{a}>0$ is learning rate. 
\end{proposition}
\begin{proof}\nonumber
At first, we differentiate the objective function without the regularization term by $\boldsymbol{A}$.
\begin{align}
    &\frac{\partial L(\boldsymbol{A},\boldsymbol{B},\boldsymbol{\varGamma}^{*})}{\partial \boldsymbol{A}}=
    \frac{\partial \left\{\|\bm{Y}-\frac{1}{2}\bm{T}^{(\mathrm{obs})}\boldsymbol{XA\varGamma}^{*}\|^{2}_{F}+\omega\|\boldsymbol{X}-\boldsymbol{XAB}^{\prime}\|^{2}_{F}\right\}}{\partial \boldsymbol{A}}
    \\
    &\quad=\left(-\boldsymbol{X}^{\prime}\boldsymbol{T}^{(\mathrm{obs})}\bm{Y\varGamma}^{*^\prime}+\frac{1}{2}\boldsymbol{X}^{\prime}\boldsymbol{XA\varGamma}^{*}\boldsymbol{\varGamma}^{*^\prime}\right)+\omega\left(-2\boldsymbol{X}^{\prime}\boldsymbol{XB}+2\boldsymbol{X}^{\prime}\boldsymbol{XAB}^{\prime}\boldsymbol{B}\right)
    \\
    &\quad=\left(-\boldsymbol{X}^{\prime}\boldsymbol{T}^{(\mathrm{obs})}\bm{Y\varGamma}^{*^\prime}+\frac{1}{2}\boldsymbol{X}^{\prime}\boldsymbol{XA\varGamma}^{*}\boldsymbol{\varGamma}^{*^\prime}\right)+\omega\left(-2\boldsymbol{X}^{\prime}\boldsymbol{XB}+2\boldsymbol{X}^{\prime}\boldsymbol{XA}\right)
\end{align}
Let $\alpha_{a}>0$ be the learning rate,
so $\boldsymbol{A}$ is updated to $\boldsymbol{A}^{\dagger}$ using the gradient descent method.
\begin{align}
    \boldsymbol{A}^{\dagger}&=\boldsymbol{A}-\alpha_{a}\left\{\left(-\boldsymbol{X}^{\prime}\boldsymbol{T}^{(\mathrm{obs})}\bm{Y\varGamma}^{*^\prime}+\frac{1}{2}\boldsymbol{X}^{\prime}\boldsymbol{XA\varGamma}^{*}\boldsymbol{\varGamma}^{*^\prime}\right)+\omega\left(-2\boldsymbol{X}^{\prime}\boldsymbol{XB}+2\boldsymbol{X}^{\prime}\boldsymbol{XA}\right)\right\}
\end{align}
Finally, we use the soft-thresholding operator (Tibshirani, 1996) for $\boldsymbol{A}^{\dagger}$. Then it agrees with Eq. $(\ref{soft})$.
\end{proof}
\begin{proposition}
Given $\boldsymbol{A}$ and $\boldsymbol{\varGamma}^{*}$, $\boldsymbol{B}$ is updated to $\widehat{\boldsymbol{B}}$ in order to minimize the objective function by the singular value decomposition of $(\boldsymbol{X}^{\prime}\boldsymbol{X})\boldsymbol{A}$.
\begin{align}
\label{BBB}
    \widehat{\boldsymbol{B}}\leftarrow \boldsymbol{UV}^{\prime}
\end{align}
where $(\boldsymbol{X}^{\prime}\boldsymbol{X})\boldsymbol{A}=\boldsymbol{U\varPsi V}^{\prime}$ by the singular value decomposition. Here, $\bm{U}\in\mathbb{R}^{(m+1)\times d}$ and $\bm{V}\in\mathbb{R}^{d\times d}$ are left- and right-singular vectors, respectively. $\bm{\varPsi}=\mathrm{diag}(\psi_1,\psi_2,\cdots,\psi_d)$ is square diagonal matrix.
\end{proposition}
\begin{proof}
This proof can be conducted in the same way as Theorem 4. in Zou et al. (2006).
\end{proof}
\begin{proposition}
Given $\boldsymbol{A}$ and $\boldsymbol{B}$, $\boldsymbol{\varGamma}^{*}$ is updated to $\widehat{\boldsymbol{\varGamma}^{*}}$ in order to minimize the objective function by the proximal gradient method.
 \begin{align}\label{soft2}
    \widehat{\gamma^{*}}_{ij}&\leftarrow \left\{
\begin{array}{ll}
\gamma^{*^\dagger}_{ij}-\lambda_{\gamma^*} & (\gamma^{*^\dagger}_{ij}>0\ and\ \lambda_{\gamma^*}<|\gamma^{*^\dagger}_{ij}|) \\
\gamma^{*^\dagger}_{ij}+\lambda_{\gamma^*}& (\gamma^{*^\dagger}_{ij}<0\ and\ \lambda_{\gamma^*}<|\gamma^{*^\dagger}_{ij}|)\\
0&(\lambda_{\gamma^*}\geq |\gamma^{*^\dagger}_{ij}|)
\end{array}
\right.,
\end{align}
where $\gamma^{*^\dagger}_{ij}$ denotes the $(i,j)$ elements of
\begin{align}\nonumber
    \boldsymbol{\varGamma}^{*^\dagger}&=\boldsymbol{\varGamma}^{*}-\alpha_{\gamma^*}\left(-\boldsymbol{A}^{\prime}\boldsymbol{X}^{\prime}\boldsymbol{T}^{(\mathrm{obs})}\bm{Y}+\frac{1}{2}\boldsymbol{A}^{\prime}\boldsymbol{X}^{\prime}\boldsymbol{XA\varGamma}^{*}\right),
\end{align}
and $\alpha_{\gamma^*}>0$ is learning rate.
\end{proposition}
\begin{proof}
When updating $\boldsymbol{\varGamma}^{*}$, it is the same as in the case of $\boldsymbol{A}$. We differentiate the objective function without the regularization term by $\boldsymbol{\varGamma}^{*}$
\begin{align}\nonumber
    \frac{\partial L(\boldsymbol{A},\boldsymbol{B},\boldsymbol{\varGamma}^{*})}{\partial \boldsymbol{\varGamma}^*}=&\frac{\partial \left\{\|\bm{Y}-\frac{1}{2}\bm{T}^{(\mathrm{obs})}\boldsymbol{XA\varGamma}^{*}\|^{2}_{F}+\omega\|\boldsymbol{X}-\boldsymbol{XAB}^{\prime}\|^{2}_{F}\right\}}{\partial \boldsymbol{\varGamma}^{*}}
    \\
    \nonumber
    =&-\boldsymbol{A}^{\prime}\boldsymbol{X}^{\prime}\boldsymbol{T}^{(\mathrm{obs})}\bm{Y}+\frac{1}{2}\boldsymbol{A}^{\prime}\boldsymbol{X}^{\prime}\boldsymbol{XA\varGamma}^{*}
\end{align}
Let $\alpha_{\gamma^*}>0$ be the learning rate, so $\boldsymbol{\varGamma}^{*}$ is updated to $\boldsymbol{\varGamma}^{*^\dagger}$ using the gradient descent method.
\begin{align}\nonumber
\boldsymbol{\varGamma}^{*^\dagger}&=\boldsymbol{\varGamma}^{*}-\alpha_{\gamma^*}\left(-\boldsymbol{A}^{\prime}\boldsymbol{X}^{\prime}\boldsymbol{T}^{(\mathrm{obs})}\bm{Y}+\frac{1}{2}\boldsymbol{A}^{\prime}\boldsymbol{X}^{\prime}\boldsymbol{XA\varGamma}^{*}\right)
\end{align}
Finally, we use the soft-thresholding operator for $\boldsymbol{\varGamma}^{*^\dagger}$. Then it agrees with Eq. $(\ref{soft2})$.
\end{proof}
\section{Objective function and algorithm of SMR-MOM for binary case}
In this section, we present the objective function of SMR-MOM for binary case and our algorithm for solving optimization problems.
In the objective function of SMR-MOM for multiple continuous outcomes, the first term is the squared loss, but when the outcome is binary, it is the negative log-likelihood. In summary, this objective function is the extension of sparse principal component logistic regression: SPCLR (Kawano et al., 2018) to the multiple outcomes.

\subsection{Objective function of SMR-MOM for binary case}\label{bbb}
When the multiple outcomes is binary, it is assumed that each pair of outcomes is independent.
Then, the log-likelihood for the parameters $\boldsymbol{A},\boldsymbol{\varGamma}^{*}$ can be described as follows by using Eq. $(\ref{mc2n})$,
\begin{align}\nonumber
    \ell_b(\boldsymbol{A},\boldsymbol{\varGamma}^{*})&=\sum_{i=1}^{n}\sum_{j=1}^{p}
    \left[y_{ij}\mathrm{log}P(Y_{ij}=1|\boldsymbol{X},\boldsymbol{T}=\boldsymbol{T}^{(\mathrm{obs})})\right.
    \\
    &\quad\left.+(1-y_{ij})\mathrm{log}\left\{1-P(Y_{ij}=1|\boldsymbol{X},\boldsymbol{T}=\boldsymbol{T}^{(\mathrm{obs})})\right\} \right]\\
    \nonumber
   &=\sum_{i=1}^{n}\sum_{j=1}^{p}
    \left[\frac{t_{i}}{2}y_{ij}\boldsymbol{\gamma}_{j}^{*^\prime}\boldsymbol{A}^\prime\boldsymbol{x}_{i}-\mathrm{log}\left\{1+\mathrm{exp}\left(\frac{t_{i}}{2} \boldsymbol{\gamma}_{j}^{*^\prime}\boldsymbol{A}^\prime\boldsymbol{x}_{i}\right)\right\} \right].
\end{align}
where $y_{ij}$ is the measured value of random variable $Y_{ij}$.
The objective function of SMR-MOM for multiple binary outcomes is formulated as follows
\begin{align}\label{bi1}
    &\underset{\boldsymbol{A},\boldsymbol{B},\boldsymbol{\varGamma}^{*}}{\mathrm{min}}\left\{-\ell_b(\boldsymbol{A},\boldsymbol{\varGamma}^{*})+\omega\|\boldsymbol{X}-\boldsymbol{XAB}^{\prime}\|^{2}_{F}+\lambda_{a}\sum_{k=1}^{d}\|\boldsymbol{a}_{k}\|_{1}+\lambda_{\gamma^*}\sum_{l=1}^{p}\|\boldsymbol{\gamma}^{*}_{l}\|_{1}\right\}
    \\\nonumber
    &\qquad\qquad\mathrm{subject\ to\ }\boldsymbol{B}^{\prime}\boldsymbol{B}=\boldsymbol{I}.
    \end{align}
    where,
    \begin{align*}
    \ell_b(\boldsymbol{A},\boldsymbol{\varGamma}^{*})=\sum_{i=1}^{n}\sum_{j=1}^{p}\frac{t_{i}}{2}y_{ij}\boldsymbol{\gamma}_{j}^{*^\prime}\boldsymbol{A}^\prime\boldsymbol{x}_{i}-\sum_{i=1}^{n}\sum_{j=1}^{p}\mathrm{log}\left\{1+\mathrm{exp}\left(\frac{t_{i}}{2} \boldsymbol{\gamma}_{j}^{*^\prime}\boldsymbol{A}^\prime\boldsymbol{x}_{i}\right)\right\},
\end{align*}
$\omega$ is a positive tuning parameter for dimensional reduction, and $\boldsymbol{B}=(\boldsymbol{b}_{1},\boldsymbol{b}_{2},\cdots,\boldsymbol{b}_{d})\in\mathbb{R}^{(m+1)\times d}$ is loading matrix for the latent variables, and $\lambda_{a},\lambda_{\gamma^*}(\geq0)$ are regularization parameters. The first term of this objective function is negative log-likelihood and the second term is loss of PCA, and the others are the $L_{1}$-regularized terms.
\subsection{Algorithm of SMR-MOM for binary case}
In this subsection, we show our algorithm for solving optimization problems and how to update parameters.
Each parameter is updated in the same way as when the outcome is continuous.
\par
\begin{algorithm}[H]\label{al2}
\SetAlgoLined
\KwResult{$\boldsymbol{A},\boldsymbol{B},\boldsymbol{\varGamma}^{*}$}
 \textit{initialize} $\boldsymbol{A},\boldsymbol{B},\boldsymbol{\varGamma}^{*}$\\
 \While{the objective function converges}{
  \textit{Update $\boldsymbol{A}$ using proximal gradient method (by Eq. (\ref{soft3}))}\\
  \textit{Update $\boldsymbol{B}$ using singular value decomposition (by Eq. (\ref{softmax}))}\\
  \textit{Update $\boldsymbol{\varGamma}^{*}$ using proximal gradient method (by Eq. (\ref{soft4}))}\\
 }
 \caption{SMR-MOM for multiple binary outcomes}
\end{algorithm}
In order to estimate these parameters, we use the following propositions.
\begin{proposition}
Given $\boldsymbol{B}$ and $\boldsymbol{\varGamma}^{*}$, the $k$-th column vector of $\boldsymbol{A}$ i.e. $\boldsymbol{a}_{k}$ is updated to $\widehat{\boldsymbol{a}}_{k}$ in order to minimize Eq. (\ref{bi1}) by proximal gradient method.
 \begin{align}\label{soft3}
    \widehat{a}_{k_{(h)}}&\leftarrow \left\{
\begin{array}{ll}
a_{k_{(h)}}^{\dagger}-\lambda_{a} & (a_{k_{(h)}}^{\dagger}>0\ and\ \lambda_{a}<|a_{k_{(h)}}^{\dagger}|) \\
a_{k_{(h)}}^{\dagger}+\lambda_{a}& (a_{k_{(h)}}^{\dagger}<0\ and\ \lambda_{a}<|a_{k_{(h)}}^{\dagger}|)\\
0&(\lambda_{a}\geq |a_{k_{(h)}}^{\dagger}|)\quad\quad\quad\quad
\end{array}
\right. \\
\nonumber
    \boldsymbol{a}_{k}^{\dagger}&=\boldsymbol{a}_{k}-\alpha_{a}\left\{-\frac{\partial \ell(\boldsymbol{A},\boldsymbol{\varGamma}^{*})}{\partial \boldsymbol{a}_{k}}+\omega(-2\boldsymbol{X}^{\prime}\boldsymbol{Xb}_{k}+2\boldsymbol{X}^{\prime}\boldsymbol{Xa}_{k})\right\}\\
    \nonumber
    \frac{\partial \ell_b(\boldsymbol{A},\boldsymbol{\varGamma}^{*})}{\partial \boldsymbol{a}_{k}}
    &=\sum_{i=1}^{n}\sum_{j=1}^{p}\frac{t_{i}}{2}y_{ij}\gamma_{j_{(k)}}^{*}\boldsymbol{x}_{i}-\sum_{i=1}^{n}\sum_{j=1}^{p}\frac{\frac{t_i}{2}\gamma_{j_{(k)}}^{*}\boldsymbol{x}_{i}\mathrm{exp}\left(t_{i}\boldsymbol{\gamma}_{j}^{*^\prime}\boldsymbol{A}^\prime\boldsymbol{x}_{i}/2\right)}{1+\mathrm{exp}\left(t_{i} \boldsymbol{\gamma}_{j}^{*^\prime}\boldsymbol{A}^\prime\boldsymbol{x}_{i}/2\right)}
\end{align}
where $\widehat{a}_{k_{(h)}}$ and $a_{k_{(h)}}^{\dagger}$ denotes the $h$-th elements of $\widehat{\boldsymbol{a}}_{k}$ and $\boldsymbol{a}_{k}^{\dagger}$, respectively. $\alpha_{a}\geq0$ is learning rate.
\end{proposition}
\begin{proof}\nonumber
We differentiate the log-likelihood by the $k$-th column vector of $\boldsymbol{A}$ i.e. $\boldsymbol{a}_{k}$.
\begin{align}
    \frac{\partial \ell_b(\boldsymbol{A},\boldsymbol{\varGamma}^{*})}{\partial \boldsymbol{a}_{k}}=\sum_{i=1}^{n}\sum_{j=1}^{p}\frac{t_{i}}{2}y_{ij}\gamma_{j_{(k)}}^{*}\boldsymbol{x}_{i}-\sum_{i=1}^{n}\sum_{j=1}^{p}\frac{\gamma_{j_{(k)}}^{*}\boldsymbol{x}_{i}\mathrm{exp}\left(t_{i} \boldsymbol{\gamma}_{j}^{*^\prime}\boldsymbol{A}^\prime\boldsymbol{x}_{i}/2\right)}{1+\mathrm{exp}\left(t_{i} \boldsymbol{\gamma}_{j}^{*^\prime}\boldsymbol{A}^\prime\boldsymbol{x}_{i}/2\right)}
\end{align}
where $\gamma_{j_{(k)}}^{*}$ denotes the $k$-th element of $\boldsymbol{\gamma}_{j}^{*}$. And then, we differentiate the loss function of PCA by $\boldsymbol{a}_{k}$.
\begin{align}
    \frac{\partial \|\boldsymbol{X}-\boldsymbol{XAB}^{\prime}\|^{2}_{F}}{\partial \boldsymbol{a}_{k}}&=\frac{\partial \mathrm{tr}(\boldsymbol{X}-\boldsymbol{XAB}^{\prime})^{\prime}(\boldsymbol{X}-\boldsymbol{XAB}^{\prime})}{\partial \boldsymbol{a}_{k}}\\
    &=-2\boldsymbol{X}^{\prime}\boldsymbol{Xb}_{k}+2\boldsymbol{X}^{\prime}\boldsymbol{Xa}_{k}
\end{align}
Let $\alpha_{a}\geq0$ be learning rate, and $\boldsymbol{a}_{k}$ is updated to $\boldsymbol{a}_{k}^{\dagger}$ using gradient descent method.
\begin{align}
\boldsymbol{a}_{k}^{\dagger}&=\boldsymbol{a}_{k}-\alpha_{a}\cdot\left\{-\frac{\partial \ell(\boldsymbol{A},\boldsymbol{\varGamma}^{*})}{\partial \boldsymbol{a}_{k}}+\omega(-2\boldsymbol{X}^{\prime}\boldsymbol{Xb}_{k}+2\boldsymbol{X}^{\prime}\boldsymbol{Xa}_{k})\right\}
\end{align}
Finally, we use the soft-thresholding operator for $\boldsymbol{a}_{k}^{\dagger}$. Then it agrees with Eq. $(\ref{soft3})$.
\end{proof}
\begin{proposition}
Given $\boldsymbol{A}$ and $\boldsymbol{\varGamma}^{*}$, $\boldsymbol{B}$ is updated to $\widehat{\boldsymbol{B}}$ in order to minimize Eq. (\ref{bi1}) by singular value decomposition of $(\boldsymbol{X}^{\prime}\boldsymbol{X})\boldsymbol{A}$.
\begin{align}\label{softmax}
    \widehat{\boldsymbol{B}}\leftarrow \boldsymbol{UV}^{\prime}
\end{align}
where $(\boldsymbol{X}^{\prime}\boldsymbol{X})\boldsymbol{A}=\boldsymbol{UDV}^{\prime}$ by singular value decomposition.
\end{proposition}
\begin{proof}
This proof can be conducted as the same way of Theorem 4. in Zou et al. (2006).
\end{proof}
\begin{proposition}
Given $\boldsymbol{A}$ and $\boldsymbol{B}$, the $j$-th column vector of $\boldsymbol{\varGamma}^{*}$ i.e. $\boldsymbol{\gamma}^{*}_{j}$ is updated to $\widehat{\boldsymbol{\gamma}^{*}_{j}}$ in order to minimize Eq. (\ref{obj}) by proximal gradient method.
 \begin{align}\label{soft4}
    \widehat{\gamma^{*}_{j}}_{(h)}&\leftarrow \left\{
\begin{array}{ll}
\gamma^{*^\dagger}_{j_{(h)}}-\lambda_{\gamma^*} & (\gamma^{*^\dagger}_{j_{(h)}}>0\ and\ \lambda_{\gamma^*}<|\gamma^{*^\dagger}_{j_{(h)}}|) \\ 
\gamma^{*^\dagger}_{j_{(h)}}+\lambda_{\gamma^*}& (\gamma^{*^\dagger}_{j_{(h)}}<0\ and\ \lambda_{\gamma^*}<|\gamma^{*^\dagger}_{j_{(h)}}|)\\ 
0&(\lambda_{\gamma^*}\geq |\gamma^{*^\dagger}_{j_{(h)}}|)\quad\quad\quad\quad
\end{array}
\right. \\ \nonumber
\boldsymbol{\gamma}_{j}^{*^\dagger}&=\boldsymbol{\gamma}_{j}^{*}-\alpha_{\gamma^*}\left\{-\frac{\partial \ell_b(\boldsymbol{A},\boldsymbol{\varGamma}^{*})}{\partial \boldsymbol{\gamma}^{*}_{j}}\right\}
    \\ \nonumber
    \frac{\partial \ell_b(\boldsymbol{A},\boldsymbol{\varGamma}^{*})}{\partial \boldsymbol{\gamma}^{*}_{j}}&=\sum_{i=1}^{n}\sum_{j=1}^{p}\frac{t_{i}}{2}y_{ij}\boldsymbol{A}^{\prime}\boldsymbol{x}_{i}-\sum_{i=1}^{n}\sum_{j=1}^{p}\frac{\frac{t_{i}}{2}\boldsymbol{A}^{\prime}\boldsymbol{x}_{i}\mathrm{exp}\left(t_{i} \boldsymbol{\gamma}_{j}^{*^\prime}\boldsymbol{A}^\prime\boldsymbol{x}_{i}/2\right)}{1+\mathrm{exp}\left(t_{i} \boldsymbol{\gamma}_{j}^{*^\prime}\boldsymbol{A}^\prime\boldsymbol{x}_{i}/2\right)}
\end{align}
where $\widehat{\gamma^{*}_{j}}_{(h)}$ and $\gamma^{*^\dagger}_{j_{(h)}}$ denotes the $h$-th elements of $\widehat{\boldsymbol{\gamma}^{*}_{j}}$ and $\boldsymbol{\gamma}_{j}^{*^\dagger}$, respectively. $\alpha_{\gamma^*}\geq0$ is learning rate.
\end{proposition}
\begin{proof}\nonumber
We differentiate the log-likelihood by the $j$-th column vector of $\boldsymbol{\varGamma}^{*}$ i.e. $\boldsymbol{\gamma}_{j}^{*}$.
\begin{align}
\frac{\partial \ell_b(\boldsymbol{A},\boldsymbol{\varGamma}^{*})}{\partial \boldsymbol{\gamma}_{j}^{*}}=\sum_{i=1}^{n}\sum_{j=1}^{p}\frac{t_{i}}{2}y_{ij}\boldsymbol{A}^{\prime}\boldsymbol{x}_{i}-\sum_{i=1}^{n}\sum_{j=1}^{p}\frac{\frac{t_{i}}{2}\boldsymbol{A}^{\prime}\boldsymbol{x}_{i}\mathrm{exp}\left(t_{i} \boldsymbol{\gamma}_{j}^{*^\prime}\boldsymbol{A}^\prime\boldsymbol{x}_{i}/2\right)}{1+\mathrm{exp}\left(t_{i} \boldsymbol{\gamma}_{j}^{*^\prime}\boldsymbol{A}^\prime\boldsymbol{x}_{i}/2\right)}
\end{align}
The loss of PCA is not related to $\boldsymbol{\gamma}_{j}^{*}$, so the value of differentiating by $\boldsymbol{\gamma}_{j}^{*}$ is zero. Let $\alpha_{\gamma^*}\geq 0$ be learning rate, and $\boldsymbol{\gamma}_{j}^{*}$ is updated to $\boldsymbol{\gamma}_{j}^{*^\dagger}$ using gradient descent method.
\begin{align}
    \boldsymbol{\gamma}_{j}^{*^\dagger}&=\boldsymbol{\gamma}_{j}^{*}-\alpha_{\gamma^*}\left\{-\frac{\partial \ell_b(\boldsymbol{A},\boldsymbol{\varGamma}^{*})}{\partial \boldsymbol{\gamma}^{*}_{j}}\right\}
\end{align}
Finally, we use the soft-thresholding operator for $\boldsymbol{\gamma}_{j}^{*^\dagger}$. Then it agrees with Eq. $(\ref{soft4})$.
\end{proof}
\section{Generalization of SMR-MOM}
In this section, we generalize SMR-MOM for GLM like as Kawano et al. (2018) approach.
This allows us to deal with various types of multiple outcomes, including count and multiclass data.
\par
 Firstly, we rewrite Eq. $(\ref{smr})$ as follows,
\begin{align}
    \bm{Y}=\frac{1}{2}\bm{T}^{(\mathrm{obs})}\bm{XA\varGamma}^*+\Tilde{\bm{E}}.
\end{align}
We assume that the multiple outcome distributed from the exponential family (see e.g. Johnson et al., 1997)
\begin{align*}
P(Y_i=\bm{y}_i|\bm{x}_i,T_i=t_i; \bm{\eta}, h, g)=h(\bm{y}_i)g(\bm{\eta})\exp \{\bm{\eta}^\prime u(\bm{y}_i)\},
\end{align*}
where  $\bm{\eta}$ is parameter, $h(\cdot)$ and $g(\cdot)$ are given function.
$u(\cdot)$ represents arbitrary function of $\bm{y}_i$.
Here, we consider the following minimization problem
\begin{align}\label{mini}
    &\underset{\boldsymbol{A},\boldsymbol{B},\bm{C},\bm{F},\boldsymbol{\varGamma}^{*}}{\mathrm{min}}\left\{L_{\mathrm{reg}}(\boldsymbol{A},\boldsymbol{\varGamma}^{*})+\omega L_{\mathrm{DR}}(\bm{C},\bm{F},\bm{B})+P_1(\bm{A})+P_2(\bm{\varGamma}^*)\right\},
\end{align}
where
\begin{gather*}
L_{\mathrm{reg}}(\boldsymbol{A},\boldsymbol{\varGamma}^{*})=-\sum_{i=1}^n \log P(Y_{i}=\bm{y}_{i}|\bm{x}_i,T_i=t_i)
\\
L_{\mathrm{DR}}(\bm{C},\bm{F},\bm{B})=\|\bm{XC}-\bm{FB}^\prime\|_F^2
\\
P_1(\bm{A})=\sum_{k=1}^d \lambda_k p_k(\|\bm{a}_k\|),
\quad 
P_2(\bm{\varGamma}^*)=\sum_{l=1}^p \lambda_{\gamma^*} p_l(\|\bm{\gamma}^*_l\|)
\end{gather*}
and parameter matrices $\bm{C},\bm{F}$ and $\bm{B}$ are appropriately constrained according to the dimensional reduction method.
$p_k(\cdot)\ (k=1,2,\cdots,d)$ and $p_l(\cdot)\ (l=1,2,\cdots,p)$ are given nonnegative penalty functions, and
$\omega$ is a positive tuning parameter, and $\lambda_{a},\lambda_{\gamma^*}(\geq0)$ are regularization parameters.
The loss function $L_{\mathrm{reg}}(\boldsymbol{A},\boldsymbol{\varGamma}^{*})$ is the negative log-likelihood, $L_{\mathrm{DR}}(\bm{C},\bm{F},\bm{B})$ is the loss function of dimensional reduction method.
$P_1(\bm{A})$ and $P_2(\bm{\varGamma}^*)$ are penalty terms of $\bm{A}$ and $\bm{\varGamma}^*$, respectively.
Here after, we call this generalized SMR-MOM as GSMR-MOM.
In SMR-MOM, PCA and $L_1$ regularized were used as a method of dimensional reduction and penalty function, respectively.
\subsection{GSMR-MOM for multiclass case}
We assume that multiple outcome $Y_i$ follows a $p$-dimensional multinomial distribution.
 The multiclass-logistic regression model is given by
\begin{align*}
    P(Y_{ij}=y_{ij}|\bm{x}_i,T_i=t_i)=
    \frac{\exp (t_i \bm{\gamma}_j^{*^\prime}\bm{A}^\prime \bm{x}_i/2)}{\sum_{j=1}^p \exp (t_i \bm{\gamma}_j^{*^\prime}\bm{A}^\prime \bm{x}_i/2)}.
\end{align*}
Here, it is assumed that each pair of outcomes is independent.
Then, $L_{\mathrm{reg}}$ described as follows
\begin{align*}
    L_{\mathrm{reg}}
    =-\sum_{i=1}^{n}\sum_{j=1}^{p}y_{ij}\left\{\frac{1}{2}t_i \bm{\gamma}_j^{*^\prime}\bm{A}^\prime \bm{x}_i-\mathrm{log}\sum_{j=1}^{p}\mathrm{exp}(t_i \bm{\gamma}_j^{*^\prime}\bm{A}^\prime \bm{x}_i/2)\right\}.
\end{align*}
\par
Since $\mathrm{log}\sum_{j=1}^{p}\mathrm{exp}(t_i \bm{\gamma}_j^{*^\prime}\bm{A}^\prime \bm{x}_i/2)$ is convex (see e.g. Boyd and Vandenberghe, 2004), Eq. $(\mathrm{\ref{mini}})$ can be optimized by the gradient method.
\subsection{GSMR-MOM for count case}
Suppose that we have a count outcome $Y_{ij}\in\{0\}\cup \mathbb{N}$, where $\mathbb{N}$ is a set of natural number. The Poisson regression model is given by
\begin{align*}
    P(Y_{ij}=y_{ij}|\bm{x}_i,T_i=t_i)
    =\frac{\exp \left(y_{ij}t_i \bm{\gamma}_j^{*^\prime}\bm{A}^\prime \bm{x}_i/2-\exp (t_i \bm{\gamma}_j^{*^\prime}\bm{A}^\prime \bm{x}_i/2)\right)}{y_{ij}!},
\end{align*}
where $!$ represents a factorial and $0!=1$.
Here, it is assumed that each pair of outcomes is independent. Then, $L_{\mathrm{reg}}$ is as follows
\begin{align*}
    L_{\mathrm{reg}}
    =-\sum_{i=1}^n\sum_{j=1}^p
    \left\{\frac{1}{2}y_{ij}t_i \bm{\gamma}_j^{*^\prime}\bm{A}^\prime \bm{x}_i-\exp (t_i \bm{\gamma}_j^{*^\prime}\bm{A}^\prime \bm{x}_i/2)-\log y_{ij}!\right\}.
\end{align*}
\par
Since the exponential function is convex, Eq. $(\mathrm{\ref{mini}})$ can be optimized by the gradient method as before.
\section{Numerical study}
In this section, we present the simulation designs and the results of numerical simulations.
Here, we consider the case where multiple outcomes are continuous or binary.
After comparison with the proposed models, here we introduce the low-rank full model.
This model is described as follows,
\par 
low-rank full model:
\begin{align*}
    \boldsymbol{Y}=g(\boldsymbol{X})+\frac{1}{2}\boldsymbol{T}^{(\mathrm{obs})}\boldsymbol{XA\varGamma}^*+\boldsymbol{E}.
\end{align*}
where $g:\mathbb{R}^{n\times (m+1)}\mapsto \mathbb{R}^{n\times p}$ is the arbitrary function for the main effect term.
From this model, when $\boldsymbol{Y}$ is continuous, the model is described as follows,
\begin{align*}
    &P(Y_{i}=\boldsymbol{y}_{i}|\boldsymbol{X},\boldsymbol{T}=\boldsymbol{T}^{(\mathrm{obs})})
        \\&=\frac{1}{\sqrt{(2\pi)^{p}|\boldsymbol{\Sigma}|}}\exp\left[-\frac{1}{2}\left\{\boldsymbol{y}_i-\left(f(\boldsymbol{x}_i)+\frac{t_i}{2}\boldsymbol{\varGamma}^{*^\prime}\boldsymbol{A}^\prime\boldsymbol{x}_i\right)\right\}^\prime \boldsymbol{\Sigma}^{-1}\left\{\boldsymbol{y}_i-\left(f(\boldsymbol{x}_i)+\frac{t_i}{2}\boldsymbol{\varGamma}^{*^\prime}\boldsymbol{A}^\prime\boldsymbol{x}_i\right)\right\}\right].
    \end{align*}
When $\boldsymbol{Y}$ is binary, the model is described as follows,
\begin{align*}
     P(Y_{ij}=1|\boldsymbol{X},\boldsymbol{T}=\boldsymbol{T}^{(\mathrm{obs})})=\frac{\mathrm{exp}(g(\boldsymbol{X})_{ij}+t_{i} \boldsymbol{\gamma}_{j}^{*^\prime}\boldsymbol{A}^\prime\boldsymbol{x}_{i}/2)}{1+\mathrm{exp}(g(\boldsymbol{X})_{ij}+t_{i}\boldsymbol{\gamma}_{j}^{*^\prime}\boldsymbol{A}^\prime\boldsymbol{x}_{i}/2)}.
\end{align*}
where $g(\boldsymbol{X})_{ij}$ denotes the $(i,j)$-th elements of $g(\boldsymbol{X})$.
SMR-MOM are compared with three other models. 
The objective functions and the algorithm of these three models are described below.
In the objective function of low-rank full (logistic) model, the framework of SPCR (SPCLR) will be used in the same way as Subsection \ref{aaa} and Subsection \ref{bbb}.
The basic loss function for SPCR is based on a combination of the lasso regression squared loss and PCA loss.
The basic loss function for SPCLR is based on a combination of the lasso logistic regression squared loss and PCA loss.
Therefore, tandem approach implies applying SPCA and Lasso regression independently for multiple continuous outcomes, and implies applying SPCA and Lasso logistic regression independently for multiple binary outcomes.
The algorithm for these simulations is shown in Table 1.
\begin{table}[ht]
\centering
\caption{Algorithms for multiple continuous outcomes and multiple binary outcomes}
  \begin{tabular}{lcc}
  \hline
     & Tandem (SPCA+Lasso regression)&Simultaneous (SPCR) \\ \hline
Full model & Algorithm 3 & Algorithm 4 \\
    MOM & Algorithm 5 & \textbf{Algorithm 1: SMR-MOM} \\ \hline
    & Tandem (SPCA+Lasso logistic regression)&Simultaneous (SPCLR) \\ \hline
    Full logistic model & Algorithm 3& Algorithm 4 \\
    MOM & Algorithm 5 & \textbf{Algorithm 2: SMLR-MOM}
    \\\hline
  \end{tabular}
  \label{tb:fugafuga}
\end{table}
The purpose of all these algorithms is to estimate the treatment effect $\boldsymbol{XA\varGamma}^*$.
Therefore, we calculate mean squared errors: MSEs between the true treatment effects and estimated treatment effects, to evaluate these methods.
This is because we selected these algorithms in the comparison.
We describe three algorithms for low-rank full model with tandem approach, low-rank full model with simultaneous approach, and modified outcome method with tandem approach, in Algorithm 3, 4, and 5, respectively.\\
    \begin{algorithm}[H]
\SetAlgoLined
\KwResult{$\boldsymbol{A},\boldsymbol{B},\boldsymbol{D},\boldsymbol{\varGamma}^{*}$}
 \textit{initialize} $\boldsymbol{A},\boldsymbol{B},\boldsymbol{D},\boldsymbol{\varGamma}^{*}$\\
  \textit{Update $\boldsymbol{A},\boldsymbol{B}$ using SPCA}\\
  \textit{Update $\boldsymbol{D},\boldsymbol{\varGamma}^{*}$ using lasso (logistic) regression}
 \caption{Full-Tandem}
\end{algorithm}
    \begin{algorithm}[H]
\SetAlgoLined
\KwResult{$\boldsymbol{A},\boldsymbol{B},\boldsymbol{D},\boldsymbol{\varGamma}^{*}$}
 \textit{initialize} $\boldsymbol{A},\boldsymbol{B},\boldsymbol{D},\boldsymbol{\varGamma}^{*}$\\
 \While{the objective function converges}{
  \textit{Update $\boldsymbol{A}$ using proximal gradient method}\\
  \textit{Update $\boldsymbol{B}$ using singular value decomposition}\\
  \textit{Update $\boldsymbol{D}$ using proximal gradient method}\\
  \textit{Update $\boldsymbol{\varGamma}^{*}$ using proximal gradient method}\\
 }
 \caption{Full-Simultaneous}
\end{algorithm}
    \begin{algorithm}[H]
\SetAlgoLined
\KwResult{$\boldsymbol{A},\boldsymbol{B},\boldsymbol{\varGamma}^{*}$}
 \textit{initialize} $\boldsymbol{A},\boldsymbol{B},\boldsymbol{\varGamma}^{*}$\\
  \textit{Update $\boldsymbol{A},\boldsymbol{B}$ using SPCA}\\
  \textit{Update $\boldsymbol{\varGamma}^{*}$ using lasso (logistic) regression}
 \caption{MOM-Tandem}\label{al5}
\end{algorithm}
\subsection{Continuous outcomes}
For multiple continuous outcomes, we generated a matrix corresponding to outcome $\boldsymbol{Y}\in\mathbb{R}^{n\times p}$ as $\boldsymbol{Y}=(\boldsymbol{XD})\odot(\boldsymbol{XD})+\frac{1}{2}\boldsymbol{TXA\varGamma}^{*}+\boldsymbol{E}$ where the covariates matrix $\boldsymbol{X}\in\mathbb{R}^{n\times m}\sim N(\boldsymbol{0}_{m},(1-\rho)\boldsymbol{I}_{m}+\rho\boldsymbol{1}_{m}\boldsymbol{1}_{m}^{\prime})$, $\boldsymbol{I}_{m}\in\mathbb{R}^{m\times m}$ is identity matrix, $\boldsymbol{E}\in\mathbb{R}^{n\times p}\sim N(\boldsymbol{0}_{p},\sigma_0^2\{(1-\xi)\boldsymbol{I}_{p}+\xi\boldsymbol{1}_{p}\boldsymbol{1}_{p}^{\prime}\})$, and $\odot$ represents Hadmard product. Here we assume that each column vector of $\boldsymbol{D}\in\mathbb{R}^{m\times p}$ is the same vector such as $\boldsymbol{D}=(\boldsymbol{\beta}^{*},\boldsymbol{\beta}^{*},\cdots,\boldsymbol{\beta}^{*})$ where $\boldsymbol{\beta}^{*}=(\beta^{*}_{1},\beta^{*}_{2},\cdots,\beta^{*}_{m+1})^{\prime}$.
We let $\sigma_{0}=\sqrt{2}$, $n=100$, $p=10$, $d=5$ and $m=49$ respectively. The $i$-th diagonal element of binary treatment indicator matrix $\boldsymbol{T}$, i.e., $T_{i}$ is distributed from Bernoulli distribution satisfying $P(T_i=1)=P(T_i=-1)=0.5$.
Here we assume that we have misidentified main effects that are nonlinear functions as linear.
We consider eight scenarios as the same manner of Tian et al. (2014) (See, Table \ref{design}).
In addition to the settings of Tian et al. (2014), we consider the case where the errors are correlated.
\begin{table}[htb]
\begin{center}
\caption{Simulation design}
\label{design}
\begin{tabular}{llccccc} \toprule
\it{ID}& $\beta_1^*$ & $\beta_j^*\; (j=4,5,\cdots,11)$& $\beta_l^*\; (l=12,13,\cdots,50)$ & $\rho$&$\xi$\\ \midrule
1 & $(\sqrt{6})^{-1}$ & $(2\sqrt{6})^{-1}$&0&0&0\\
2 &$(\sqrt{3})^{-1}$ &$(2\sqrt{3})^{-1}$&0&0&0\\
3 & $(\sqrt{6})^{-1}$ & $(2\sqrt{6})^{-1}$&0 &1/3&0\\
4 & $(\sqrt{3})^{-1}$ & $(2\sqrt{3})^{-1}$&0&1/3&0&
\\
5 & $(\sqrt{6})^{-1}$ & $(2\sqrt{6})^{-1}$&0&0&1/3
\\
6 & $(\sqrt{3})^{-1}$ & $(2\sqrt{3})^{-1}$&0&0&1/3\\
7 & $(\sqrt{6})^{-1}$ & $(2\sqrt{6})^{-1}$&0&1/3&1/3
\\
8 & $(\sqrt{3})^{-1}$& $(2\sqrt{3})^{-1}$&0&1/3&1/3
\\
\bottomrule
\end{tabular}
\end{center}
\end{table}
\par 
Next, we set the true value of each parameter $\boldsymbol{A}\in\mathbb{R}^{50\times 5}$ and $\boldsymbol{\varGamma}^{*}\in\mathbb{R}^{5\times 10}$ as $\boldsymbol{A}^{(\mathrm{true})}$ and $\boldsymbol{\varGamma}^{*(\mathrm{true})}$, respectively.
That is, $\boldsymbol{A}^{(\mathrm{true})}=\mathrm{b}$-$\mathrm{diag}(\boldsymbol{a},\boldsymbol{a},\boldsymbol{a},\boldsymbol{a},\boldsymbol{a})$ and $\boldsymbol{\varGamma}^{*(\mathrm{true})}=\mathrm{b}$-$\mathrm{diag}(\boldsymbol{\gamma}^*,\boldsymbol{\gamma}^*,\boldsymbol{\gamma}^*,\boldsymbol{\gamma}^*,\boldsymbol{\gamma}^*)^\prime$, where $\mathrm{b}$-$\mathrm{diag}$ represents block diagonal matrix, and $\boldsymbol{a}=(1/\sqrt{10},1/\sqrt{10},\cdots,1/\sqrt{10})^{\prime}$ and $\boldsymbol{\gamma}^{*}=(0.8,-0.8)^{\prime}$.
\par
Table $\ref{lowcon}$ gives the median and interquartile range (IQR) of MSEs for the multiple continuous outcomes.
Here, MSE is defined as $\|\boldsymbol{XA\varGamma}^*-\boldsymbol{XA}^{(\mathrm{true})}\boldsymbol{\varGamma}^{*(\mathrm{true})}\|_F^2/100$, where $\boldsymbol{A},\boldsymbol{\varGamma}^*$ are estimated values.
\begin{table}[ht]
\begin{center}
\caption{Results for continous outcomes}
\label{lowcon}
\begin{tabular}{llcccc} \toprule
\it{Model}&\multicolumn{2}{c}{\it{Setting1}}&\multicolumn{2}{c}{\it{Setting2}}\\ \cmidrule(r){2-3}
 \cmidrule(r){4-5}
 & \it{Median} & \it{IQR}& \it{Median} & \it{IQR}\\ \midrule
1 Full-Tandem & 1.082 & [1.048,\;1.122]&1.083&[1.035,\;1.131]\\
2 Full-Simultaneous&1.062 &[1.018,\;1.096]&1.067&[1.016,\;1.100]\\
3 MOM-Tandem & 1.250 & [1.181,\;1.345]&1.245 &[1.175,\;1.344]\\
4 SMR-MOM & \textbf{1.060} & [1.019,\;1.095]&\textbf{1.059}&[1.011,\;1.090]&\\ \toprule
\it{Model}&\multicolumn{2}{c}{\it{Setting3}}&\multicolumn{2}{c}{\it{Setting4}}\\ \cmidrule(r){2-3}
 \cmidrule(r){4-5}
 & \it{Median} & \it{IQR}& \it{Median} & \it{IQR}\\ \midrule
1 Full-Tandem & 1.518 & [0.863,\;2.192]&3.162&[1.847,\;4.951]\\
2 Full-Simultaneous& 1.083 &[0.972,\;1.229]&1.604&[1.336,\;1.762]\\
3 MOM-Tandem & 2.581  & [1.837,\;3.764]&2.736 &[1.801,\;5.165]\\
4 SMR-MOM & \textbf{0.803} & [0.749,\;0.874]&\textbf{1.256}&[1.018,\;1.498]&\\
\toprule \it{Model}&\multicolumn{2}{c}{\it{Setting5}}&\multicolumn{2}{c}{\it{Setting6}}\\ \cmidrule(r){2-3}
 \cmidrule(r){4-5}
 & \it{Median} & \it{IQR}& \it{Median} & \it{IQR}\\ \midrule
1 Full-Tandem & 0.611 & [0.582,\;0.633]&0.624&[0.595,\;0.649]\\
2 Full-Simultaneous&0.595 &[0.572,\;0.612]&0.601&[0.578,\;0.625]\\
3 MOM-Tandem & 0.668 & [0.643,\;0.703]&0.674 &[0.653,\;0.717]\\
4 SMR-MOM & \textbf{0.590} & [0.570,\;0.614]&\textbf{0.596}&[0.576,\;0.623]&
\\
\toprule \it{Model}&\multicolumn{2}{c}{\it{Setting7}}&\multicolumn{2}{c}{\it{Setting8}}\\ \cmidrule(r){2-3}
 \cmidrule(r){4-5}
 & \it{Median} & \it{IQR}& \it{Median} & \it{IQR}\\ \midrule
1 Full-Tandem & 1.319 & [1.137,\;1.622]&2.753&[1.545,\;5.468]\\
2 Full-Simultaneous&1.176 &[1.086,\;1.363]&1.436&[1.219,\;1.624]\\
3 MOM-Tandem & 3.420 & [2.626,\;4.285]&5.910 &[4.611,\;7.617]\\
4 SMR-MOM & \textbf{0.973} & [0.808,\;1.118]&\textbf{1.139}&[0.983,\;1.394]&
\\
\bottomrule
\end{tabular}
\end{center}
\end{table}
\par
SMR-MOM is superior to any other method in all scenarios.
The results of simultaneous estimation such as Algorithm 1 and Algorithm 4 are superior to those of tandem approaches such as Algorithm 3 and Algorithm 5, respectively.
The proposed method is the most robust against increasing correlations.
\subsection{Binary outcomes}
For multiple binary outcomes, we used the simulation design in the same manner as used for continuous outcomes.
We generated $Y_{ij}^{*}$ denoting the $(i,j)$ element of binary matrix corresponding to outcome $\boldsymbol{Y}^{*}\in\mathbb{R}^{n\times p}$ as $Y^{*}_{ij}=\mathbb{I}(Y_{ij}>0)$, where $Y_{ij}$ is the $(i,j)$ element of $\boldsymbol{Y}=(\boldsymbol{XD})\odot(\boldsymbol{XD})+\frac{1}{2}\boldsymbol{TXA\varGamma}^{*}+\boldsymbol{E}$ and $\mathbb{I}(\cdot)$ is indicator function. 
Table $\ref{lowbi}$ gives the median and IQR of MSEs for the multiple binary outcomes.
Here, MSE is defined as same as continuous case.
\begin{table}[ht]
\begin{center}
\caption{Results for binary outcomes}
\label{lowbi}
\begin{tabular}{llcccc} \toprule
\it{Model}&\multicolumn{2}{c}{\it{Setting1}}&\multicolumn{2}{c}{\it{Setting2}}\\ \cmidrule(r){2-3}
 \cmidrule(r){4-5}
 & \it{Median} & \it{IQR}& \it{Median} & \it{IQR} \\ \midrule
1 Full-Tandem & 0.672 & [0.649,\;0.697]&0.662&[0.642,\;0.693]\\
2 Full-Simultaneous& 0.630 &[0.629,\;0.631] &0.631&[0.630,\;0.633]\\
3 MOM-Tandem &0.794   &[0.632,\;0.971] &0.703&[0.631,\;0.843]\\
4 SMLR-MOM & \textbf{0.609} &[0.586,\;0.646] &\textbf{0.605}&[0.583,\;0.636]\\
\toprule
\it{Model}&\multicolumn{2}{c}{\it{Setting3}}&\multicolumn{2}{c}{\it{Setting4}} \\ \cmidrule(r){2-3}
 \cmidrule(r){4-5}
 & \it{Median} & \it{IQR}& \it{Median} & \it{IQR} \\ \midrule
1 Full-Tandem&2.732&[2.660,\;2.785]&2.654&[2.623,\;2.694]\\
2 Full-Simultaneous&2.593&[2.584,\;2.600]&2.593&[2.584,\;2.600]\\
3 MOM-Tandem &3.431&[3.181,\;3.743]&2.911&[2.668,\;3.190]\\
4 SMLR-MOM &\textbf{1.634} &[1.427,\;1.855]&\textbf{2.001}&[1.844,\;2.237]\\
\toprule \it{Model}&\multicolumn{2}{c}{\it{Setting5}}&\multicolumn{2}{c}{\it{Setting6}}\\ \cmidrule(r){2-3}
 \cmidrule(r){4-5}
 & \it{Median} & \it{IQR}& \it{Median} & \it{IQR}\\ \midrule
1 Full-Tandem & 0.839 & [0.747,\;0.954]&0.860&[0.743,\;0.999]\\
2 Full-Simultaneous&0.609 &[0.587,\;0.649]&0.611&[0.590,\;0.645]\\
3 MOM-Tandem & 0.746 & [0.696,\;0.817]&0.762 &[0.693,\;0.822]\\
4 SMR-MOM & \textbf{0.584} & [0.564,\;0.614]&\textbf{0.587}&[0.563,\;0.621]&
\\
\toprule \it{Model}&\multicolumn{2}{c}{\it{Setting7}}&\multicolumn{2}{c}{\it{Setting8}}\\ \cmidrule(r){2-3}
 \cmidrule(r){4-5}
 & \it{Median} & \it{IQR}& \it{Median} & \it{IQR}\\ \midrule
1 Full-Tandem & \textbf{1.149} & [0.933,\;1.377]&\textbf{1.726}&[1.350,\;2.065]\\
2 Full-Simultaneous&2.161 &[1.906,\;2.321]&2.307&[2.141,\;2.548]\\
3 MOM-Tandem & 2.657 & [2.323,\;2.908]&2.626 &[2.338,\;2.914]\\
4 SMR-MOM & 1.926 & [1.738,\;2.083]&2.102&[1.910,\;2.312]&
\\\bottomrule
\end{tabular}
\end{center}
\end{table}
\par
The results were similar to the interpretation for continuous outcomes when the errors were uncorrelated.
Specifically, SMLR-MOM is superior to any other method in all scenarios where there is no correlation between the errors.
Here, we discuss the results when errors are correlated.
See the Settings 5 and 6. SMLR-MOM is superior to any other method.
However, see the Settings 7 and 8, Full-tandem is superior to any other method, because the parameter update formula was derived assuming that the errors were uncorrelated. SMLR-MOM is superior to Full-Simultaneous and MOM-Tandem.
When the results of SMLR-MOM is compared to those of MOM-Tandem, those of SMLR-MOM tend to be well in all scenarios.
\section{Applications}
In this section, we applied the proposed model to real data.
We apply the proposed method to both multiple continuous outcomes and multiple binary outcomes.
After that, we identify the subgroups and interpret them.
\subsection{Data: ACTG175}
In this subsection, we explain the data \textit{AIDS clinical trials group study 175: ACTG175}.
This dataset is available in the package \textbf{speff2trial} for R.
\textit{ACTG175} was a randomized clinical trial composed of four arms: 1. zidovudine, 2. zidovudine and didanosine, 3. zidovudine and zalcitabine, and 4. didanosine, in adults infected with the human immunodeficiency virus type I whose CD4 T cell counts were between 200 and 500 per cubic millimeter (Hammer et al., 1996).\par
Here we define the test therapy as combination therapy with zidovudine and didanosine ($n$=332), and the control therapy as monotherapy with zidovudine ($n$=318). 
A persistently low CD4/CD8 ratio, despite effective antiretroviral therapy, is associated with higher morbidity of AIDS and mortality (Serrano-Villar et al., 2014).
Therefore, we take three variables as continous outcomes cd420, cd496 and cd820. And we take binary outcomes cd420\_c, cd496\_c and cd820\_c.
Each of these is a dichotomization of the outcomes used in the continuous case, with the median as the cutoff value.
We set fifteen covariates: age, wtkg, hemo, homo, drugs, karnof, oprior, z30, zprior, race, gender, str2, symptom, cd40, cd80. 
\par
This application aims to discover covariates that can be used to make subgroups.
This can be done by visualizing the relationship between the covariates, the estimated latent variables, and the outcomes using path diagrams.
\subsection{Continuous outcomes}\label{ac1}
\begin{figure}[h]
\begin{center}
\includegraphics[width=14cm]{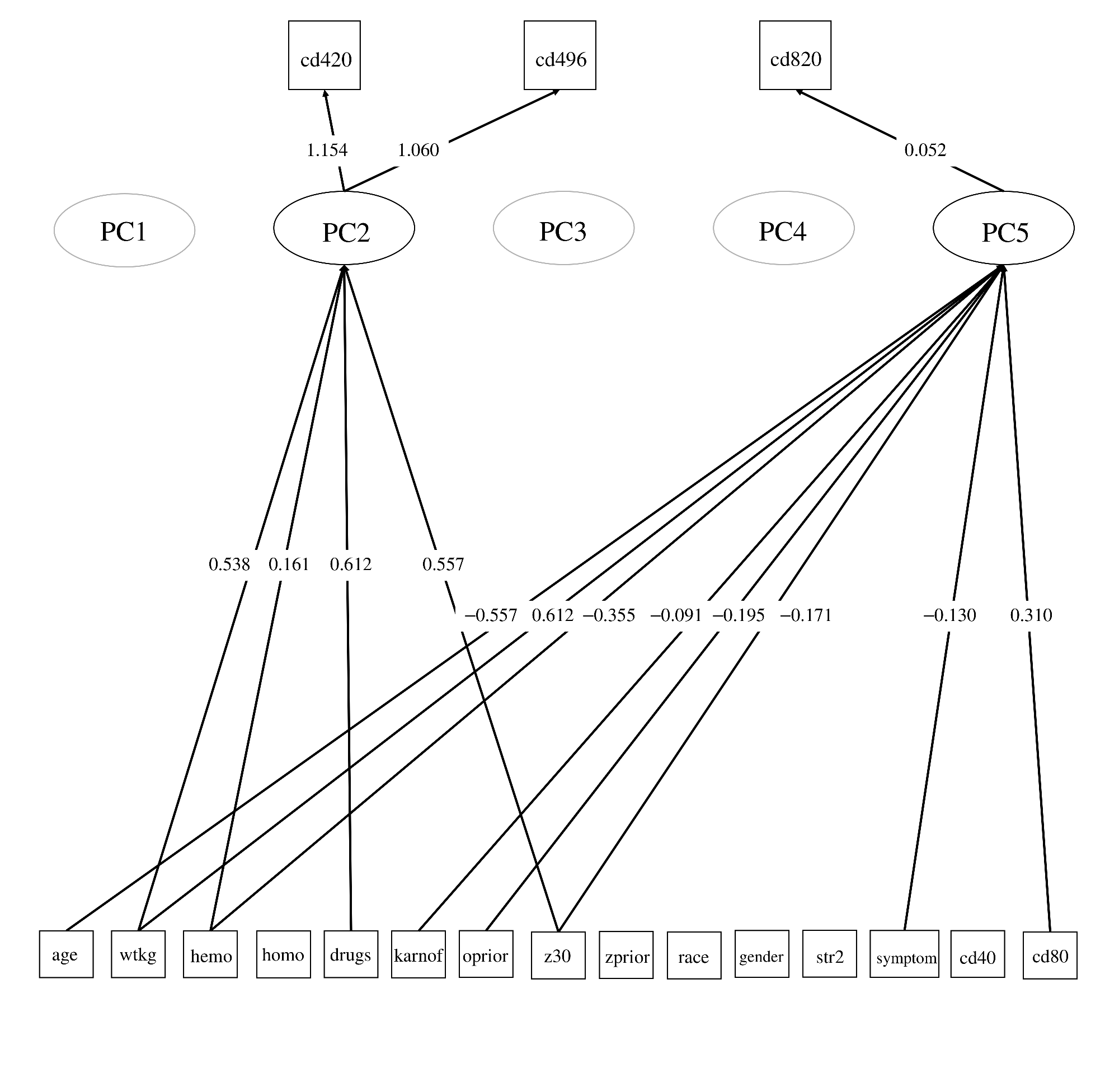}
\caption{Path diagram of SMR-MOM}
\label{ccccc}
\end{center}
\end{figure}

In this subsection, we applied the SMR-MOM to \textit{ACTG175}.
Here we assume that the number of latent variables is five.
The tuning parameter $\omega$ is set to $0.1$, $\lambda_{a}$ is selected by five-fold cross-validation from the candidate value of $(0.10,0.15,0.20,0.25,0.30)$, and $\lambda_{\gamma^*}$ is selected by cross validation.
The visualization of the estimated $\boldsymbol{A}$ and $\boldsymbol{\varGamma}^{*}$ is shown in Fig. \ref{ccccc}.
For the visualization, all coefficients that are not estimated to be zero are listed.
Two latent variables, PC2 and PC5, were identified through the SMR-MOM. 
PC2 is a component such that combination therapy has a greater effect on CD4 cells.
PC5 is a component such that combination therapy has a greater effect on CD8 cells.
For the visualization, all coefficients of $\boldsymbol{A}$ that are not estimated to be zero are listed.
Here, we describe the covariates that contribute to these components.
``$+$" means positive coefficients and ``$-$" means negative coefficients.
\\
\textbf{PC2($+$)}``wtkg", ``hemo", ``drugs", ``z30" ;\quad\textbf{PC2($-$)}\; None\\
\textbf{PC5($+$)}
``wtkg", ``cd80";
\quad\textbf{PC5($-$)}\;
``age", ``hemo", ``karnof", ``oprior", ``z30", ``symptom"
\par
From Fig. \ref{ccccc}, we found the subgroup in which test therapy was more effective in treatment effect of CD4 cells compared to control therapy.
The characteristics of the participants in the subgroup are described as follows:\\
    $\bullet$\ participants with obesity
    \ $\bullet$\ participants with zidovudine use in the 30 days prior to treatment initiation
    \ $\bullet$\ participants with hemophilia
    \ $\bullet$\ participants with history of intravenous drug use\\
In the same way, we found the following subgroups in which test therapy was more effective in treatment effect of CD8 cells compared to control therapy:
\\
    $\bullet$\ participants with obesity
    \ $\bullet$\ participants with large CD8 T cell count at baseline
\subsection{Binary outcomes}
\begin{figure}[h]
\begin{center}
\includegraphics[width=14cm]{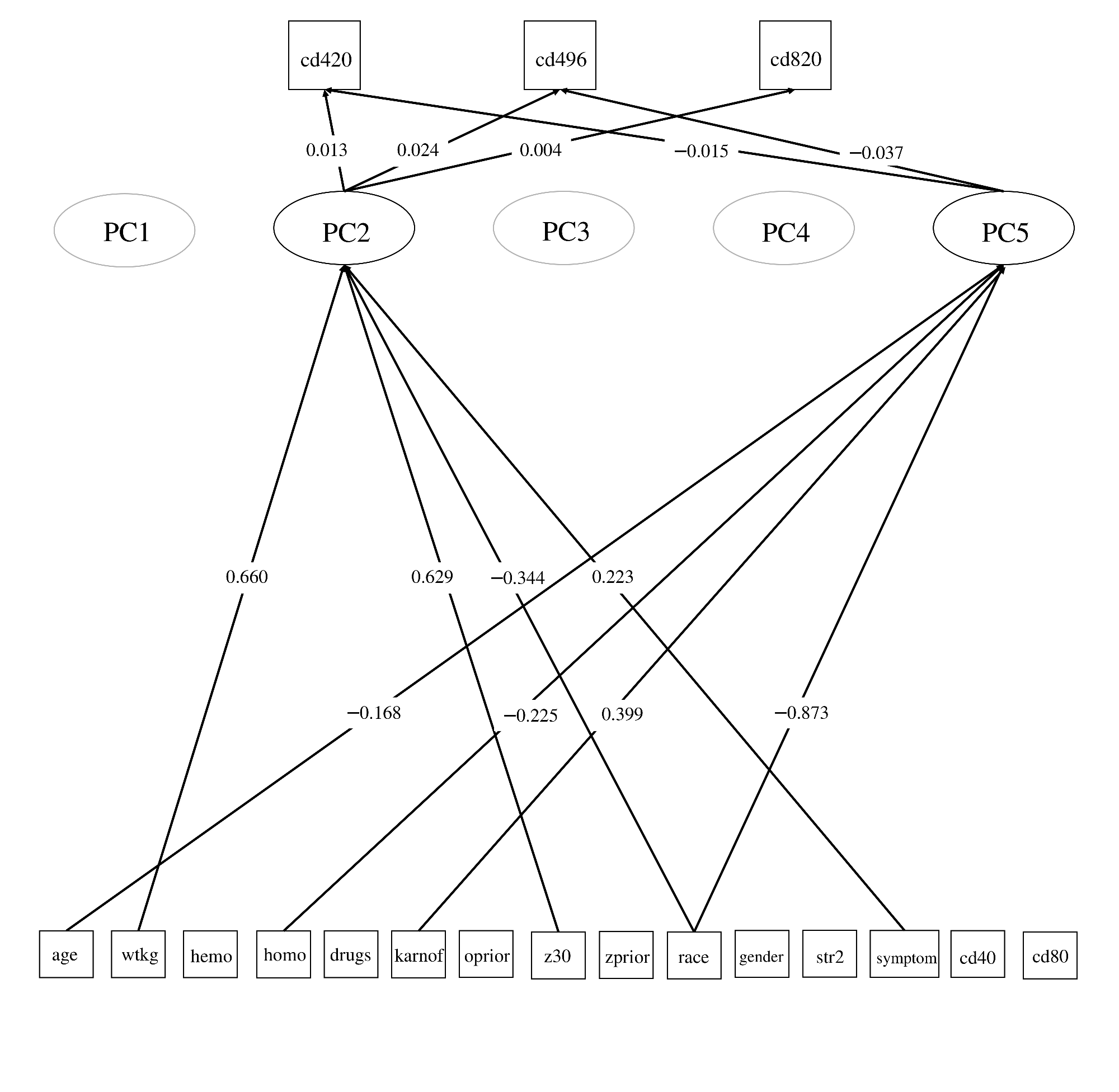}
\caption{Path diagram of SMLR-MOM}
\label{ddddd}
\end{center}
\end{figure}

In this subsection, we applied the SMR-MOM to \textit{ACTG175}.
In the same way, we assume that the number of latent variables is five.
All tuning parameters are set in the same way as continuous case. 
The visualization of the estimated $\boldsymbol{A}$ and $\boldsymbol{\varGamma}^{*}$ is shown in Fig. \ref{ddddd}.
Two latent variables, PC2 and PC5, were identified through the SMR-MOM. 
PC2 is a component such that test therapy has a greater effect on CD4 and CD8 cells compared to control therapy.
PC5 is a component such that control therapy has a greater effect on CD4 cells.
For the visualization, all coefficients of $\boldsymbol{A}$ that are not estimated to be zero are listed.
Here, we describe the covariates that contribute to these components.
\\
\textbf{PC2($+$)}``wtkg", ``z30", ``symptom";\quad
\textbf{PC2($-$)}``race"\\
\textbf{PC5($+$)}
``karnof";\quad
\textbf{PC5($-$)}
``age", ``homo", ``race"
\par
From Fig. \ref{ddddd}, we found the subgroup in which test therapy was more effective in treatment effect of CD4 cells compared to control therapy.
The characteristics of the participants in the subgroup are described as follows: 
\\
$\bullet$\ participants with obesity
    $\bullet$\ participants with zidovudine use in the 30 days prior to treatment initiation
    \\ $\bullet$\ participants with symptomatic
    \ $\bullet$\ participants with older
    \\ $\bullet$\ participants with homosexual activity
    \ $\bullet$\ participants with non-white
\section{Conclusion}
The estimation method of treatment effects for multiple outcomes has not been proposed, even though the data have been collected in clinical trials. In this paper, we propose a method to estimate the treatment effect for multiple outcomes, and confirm the usefulness of the method with numerical simulations and real data.
We proposed a method that enables the estimation of treatment effects for multiple continuous and multiple binary outcomes and the discovery of components that have a common and significant impact on treatment effects.
We called this SMR-MOM. The estimation algorithms for these proposed methods are obtained based on the proximal gradient method and singular value decomposition.
We also proposed the framework of estimating various types of outcomes as including of SMR-MOM, called GSMR-MOM. 
From the numerical study, it was found that the method that estimates the parameters by simultaneous approach has better accuracy in terms of MSEs than these by tandem approach.
In addition, from the visualization in application, we can see that these proposed methods provide ease of interpretation.
\par
In this paper, we do not select the number of principal components.
This selection can be selected by cross-validation.
In addition, this method cannot be applied to observational studies because it was proposed for randomized controlled trials. In the future, we would like to extend the method to observational studies by weighting the data using propensity scores (Rosenbaum and Rubin, 1983).
The criterion for cross-validation in numerical simulations has been "minimization of MSE", but we would like to consider making it "identification of subgroups" in the future.
The limitation of this method is that MOM is used as an efficiency augmentation in Tian et al. (2014) to provide a valid estimation of the interactions, but further investigation is needed for GLM. In the binary case, the simulation results suggest that it may be used.








\end{document}